\documentclass[epj,nopacs]{svjour}
\usepackage{graphicx}
\usepackage{amsmath}
\usepackage{amssymb}
\usepackage{booktabs}
\RequirePackage[numbers,sort&compress]{natbib}
\RequirePackage[colorlinks,citecolor=blue,urlcolor=blue,linkcolor=blue]{hyperref}
\usepackage[british]{babel}
\newcommand\as{\ensuremath{\alpha_S}}
\newcommand\asmz{\ensuremath{\alpha_S(m_Z)}}
\newcommand\tL{\ensuremath{L}}
\newcommand*{\qt}{\ensuremath{p_{\mathrm{T}}}}
\newcommand*{\pt}{\ensuremath{p_{\mathrm{T}}}}
\newcommand*{\Vboson}{\ensuremath{Z}}

\begin{document}
\title{Determination of the strong-coupling constant from the $Z$-boson transverse-momentum distribution}
\author{Stefano Camarda\inst{1} \and Giancarlo Ferrera\inst{2} \and Matthias Schott\inst{3}}
\institute{CERN, CH-1211 Geneva, Switzerland \and Dipartimento di Fisica, Universit\`a di Milano and\\ INFN, Sezione di Milano,
I-20133 Milan, Italy \and Johannes Gutenberg-Universitat Mainz (JGU),\\ Saarstr.\ 21, 55122 Mainz, Germany}
\date{}
\abstract{
  The strong-coupling constant is determined from the
  low-momentum region of the transverse-momentum distribution of $Z$ bosons
  produced through the Drell-Yan process, using predictions at third 
  order in perturbative QCD. The analysis employs a
  measurement performed in proton-antiproton collisions at a
  centre-of-mass energy of $\sqrt{s} = 1.96$~TeV with the CDF
  experiment. The determined value of the strong coupling at the
  reference scale corresponding to the $Z$-boson mass is $\asmz
  = 0.1191^{+0.0013}_{-0.0016}$.
  }

\titlerunning{\asmz{} from $Z$-boson \pt}
\maketitle
\section*{Introduction}
The coupling constant of the strong interaction is one of the fundamental
parameters of the standard model, and is the least precisely known
among the fundamental couplings in nature.
The most recent world average of the strong-coupling constant at the
scale of the $Z$-boson mass yields $\asmz = 0.1179 \pm 0.0009$,
with a relative uncertainty of $0.8\%$~\cite{ParticleDataGroup:2020ssz}.
Various different determinations contribute to the world average, and
are categorised according to their methodological approach~\cite{Salam:2017qdl}. The most
precise determinations of $\asmz$ are
from lattice QCD, with a result of $\asmz = 0.1184 \pm
0.0008$~\cite{Aoki:2021kgd}, and hadronic tau decays, with a result of $\asmz =
0.1177 \pm 0.0019$~\cite{ParticleDataGroup:2020ssz}. Tensions exist between some of the most
precise determinations of $\asmz$. For instance, several determinations
from deep inelastic lepton-nucleon scattering~\cite{Blumlein:2006be,Jimenez-Delgado:2014twa,Alekhin:2017kpj} and from
hadronic final states of electron-positron
annihilation~\cite{Abbate:2010xh,Gehrmann:2012sc,Hoang:2015hka} are
significantly lower than the lattice QCD determination.
Some of these determinations are performed at
next-to-next-to-next-to-leading order (N$^3$LO) in QCD, namely from hadronic
tau decays and low $Q^2$ con\-tin\-u\-um~\cite{Baikov:2008jh}, from
non-singlet structure functions in deep inelastic
scattering~\cite{Blumlein:2006be}, from heavy quarkonia
decays~\cite{Mateu:2017hlz,Peset:2018ria}, and from the global fit of
the electroweak observables~\cite{Haller:2018nnx,dEnterria:2020cpv}.
At hadron colliders, the strong-coupling constant has been determined
in final states with jets~\cite{CMS:2021yzl,ATLAS:2017qir}
from inclusive top quark pairs production~\cite{CMS:2018fks,CMS:2014rml,Klijnsma:2017eqp}, and
more recently from inclusive $W$ and $Z$ bosons
production~\cite{dEnterria:2019aat}.
The high-momentum region of the $Z$-boson transverse-momentum (\pt)
distribution measured at the
LHC~\cite{ATLAS:2014alx,ATLAS:2015iiu,CMS:2015hyl} was
included in parton distribution function (PDF) determinations~\cite{Boughezal:2017nla}, and contributed to the
simultaneous determination of PDFs and strong-coupling constant in Refs.~\cite{Ball:2018iqk,Hou:2019efy,Cridge:2021qfd}.
Some of these determinations, in particular those with jets in the final
state, allow probing the strong coupling at high values of
momentum transfer. 

In this context, it is highly desirable to perform alternative
determinations of $\asmz$ based on new observables and high-order theory
predictions, which can help improving the precision in the
determination of the strong coupling and resolving existing tensions.
This paper presents a new methodology for a precise determination of
$\asmz$ at hadron colliders from a semi-inclusive
(i.e.\ radiation inhibited) observable, namely the low-momentum
Sudakov~\cite{Sudakov:1954sw} region of the transverse-momentum
distribution of $Z$ bosons produced through the Drell-Yan
process~\cite{PhysRevLett.25.902.2}.
The strong force is responsible for the recoil of the $Z$ bosons,
which acquire non-zero transverse momentum from QCD radiation off the
initial-state partons, and from non-perturbative intrinsic $k_T$ effects.
The hardness of the transverse-momentum distribution is a measure of
the strength of the recoil of the $Z$ bosons, which in turn is
proportional to the strong coupling.
Compared to other determinations of $\asmz$
at hadron colliders based on either exclusive or
inclusive observables, this determination gathers all desirable
features for a precise determination: large observable sensitivity to
$\asmz$ compared to the experimental precision, high
perturbative accuracy of the theoretical prediction~\cite{Camarda:2021ict,Re:2021con,Ju:2021lah,Chen:2022cgv,Neumann:2022lft}, and
in-situ controllable non-perturbative QCD
effects~\cite{Collins:1984kg,Davies:1984sp,Ladinsky:1993zn,Ellis:1997sc,Landry:1999an,Qiu:2000hf,Kulesza:2002rh,Konychev:2005iy,Guzzi:2013aja,Collins:2014jpa,Wei:2020glg}.
The proposed methodology can be applied to proton-antiproton and
proton-proton colliders.
In this paper we consider proton-antiproton collisions data from the Tevatron
collider, because the Drell-Yan process has reduced contribution from
heavy-flavour-initiated production, compared to the proton-proton
collisions of the LHC. The application to proton-proton collisions can
profit from the large high-quality datasets already collected at the LHC
experiments, which will be further increased in the future, but could
require a more careful study of heavy-flavour-initiated production, and
is left to future work.

\section*{Methodology}
The experimental data used in the analysis is the $Z$-boson
transverse-momentum distribution measured with the CDF detector at a
centre-of-mass energy of $\sqrt{s} = 1.96$~TeV
with $2.1$~fb$^{-1}$ of integrated luminosity~\cite{CDF:2012brb}.
The measurement was performed in the electron decay channel, and
extrapolated to a kinematic region without requirements on the
transverse-momentum and pseudorapidity of the electrons. The
extrapolation to full-lepton phase space, which was based on the
measured decay lepton angular distributions~\cite{CDF:2011ksg} to avoid significant theoretical uncertainties,
enables the usage of fast analytic predictions.
In the low-momentum region below $25$~GeV, the measurement was performed in bins of
$Z$-boson transverse momentum of $0.5$~GeV. The electron resolution for
electrons of transverse momentum of $45$~GeV was approximately 1~GeV in
the central region $|\eta_e| < 1.05$, and $1.5$~GeV in the forward
region $1.2 < |\eta_e| < 2.8$, enabling small
bin-to-bin correlations at the level of $30\%$ for neighbouring bins.

The theoretical predictions are computed with the public numerical program
\texttt{DYTurbo}~\cite{Camarda:2019zyx}, which implements the resummation of logarithmically-enhanced
contributions in the small-\qt{} region of the leptons pairs at
next\--to\--next\--to\--next\--to\--leading\--logarithmic (N$^3$LL) accuracy,
combined with the hard-collinear contributions at
N$^3$LO in powers of the QCD
coupling~\cite{Camarda:2021ict}. We briefly review the resummation
formalism implemented in \texttt{DYTurbo} and developed in
Refs.~\cite{Bozzi:2005wk,Bozzi:2010xn,Catani:2015vma}.
The transverse-momentum resummed cross section for
$Z$-boson\footnote{The contribution from  $\gamma^{*}$ and
  its interference with the $Z$ boson are included throughout the
  calculation.} production can be written as
\begin{eqnarray}
\label{eq:rescross_1}
\textrm{d}\sigma^{\textrm{V}}&=&\textrm{d}\sigma^{\textrm{res}}
-\textrm{d}\sigma^{\textrm{asy}}
+\textrm{d}\sigma^{\textrm{f.o.}}\, ,
\end{eqnarray}
\noindent  where $\textrm{d}\sigma^{\textrm{res}}$ is the resummed
component of the cross-section, $\textrm{d}\sigma^{\textrm{asy}}$ is
the asymptotic term that represents the fixed-order expansion of
$\textrm{d}\sigma^{\textrm{res}}$, and
$\textrm{d}\sigma^{\textrm{f.o.}}$ is the $Z$+jet finite-order
cross section integrated over final-state QCD radiation. All the cross
sections are differential in $\qt^2$.
The resummed component $\textrm{d}\sigma^{\textrm{res}}$ is the most
important term at small \qt{} (i.e.\ $\pt \ll m_Z$). The finite-order term
$\textrm{d}\sigma^{\textrm{f.o.}}$ gives the larger net contribution
at large \qt{} (i.e.\ $\pt \sim m_Z$). The fixed-order expansion of the resummed component
$\textrm{d}\sigma^{\textrm{asy}}$ embodies the singular behaviour of
the finite-order term, providing 
a smooth behaviour of Eq.~(\ref{eq:rescross_1}) as
\qt{} approaches zero. 
The resummed component is given by\footnote{The convolution
with PDFs and the sum over different initial-state partonic
contributions are implied in the shorthand notation of
Eq.~(\ref{eq:rescross_2}).}
\begin{eqnarray}
\label{eq:rescross_2}
\textrm{d}\sigma^{\textrm{res}} &=&
\textrm{d}{\hat \sigma}^{\textrm{V}}_{\textrm{LO}}
\times \mathcal{H}_{\textrm{V}}
\times \exp\{\mathcal{G}\}
\times S_{\textrm{NP}}\;.
\end{eqnarray}
\noindent The term $\textrm{d}{\hat \sigma}^{\textrm{V}}_{\textrm{LO}}$ 
is the leading-order (LO) cross
section.

The function ${\cal H}_V$~\cite{Catani:2000vq,Catani:2013tia}
includes
the hard-collinear contributions
and it can be expanded in powers of \as{} as
\begin{equation}
\label{eq:hexpan}
{\cal H}_V(\as)=
1+ \frac{\as}{\pi} \,{\cal H}_V^{(1)} 
+ \left(\frac{\as}{\pi}\right)^2 
\,{\cal H}_V^{(2)}
+ \left(\frac{\as}{\pi}\right)^3 
\,{\cal H}_V^{(3)}+\dots \;.
\end{equation}

The universal (process independent) form factor $\exp\{{\cal G}\}$
contains all
the terms that order-by-order in $\as$ are logarithmically divergent 
as $\qt \to 0$. 
The resummed logarithmic expansion of ${\cal G}$ reads
\begin{eqnarray}
\label{eq:exponent}
{\cal G}(\as,\tL)=
\nonumber  \tL\, g^{(1)}(\as \tL)
+g^{(2)}(\as \tL) \\
+\frac{\as}{\pi} \;g^{(3)}(\as \tL)+
\left(\frac{\as}{\pi}\right)^2 \;g^{(4)}(\as \tL)+\dots\,,
\end{eqnarray}
where \tL{} is the logarithmic expansion parameter, the functions $g^{(n)}$ control and resum the $\as^k\tL^{k}$ (with $k\geq 1$) logarithmic terms in the exponent of Eq.~(\ref{eq:rescross_2})
due to soft and collinear radiation. 

The function ${\cal G}$ is singular in the region of transverse-momenta
of the order of the scale of the QCD coupling
$\Lambda_{\textrm{QCD}}$. This signals that a truly non-perturbative region is approached and 
perturbative results are not reliable.
The singular behaviour of the perturbative form factor is removed
by using the so-called $b_*$~\cite{Collins:1981va,Collins:1984kg}
regularisation procedure, in which the dependence of $\exp\{{\cal
  G}\}$ on the impact parameter $b$, that is the Fourier-conjugate variable to \qt,
is frozen before reaching the singular point by performing the replacement $b^2 \to b_*^2 = b^2
b_{\textrm{lim}}^2/( b^2 + b_{\textrm{lim}}^2)$. In the calculation
the default value of $b_{\textrm{lim}} = 3$~GeV$^{-1}$ is used.
The minimal prescription~\cite{Catani:1996yz,Laenen:2000de,Kulesza:2002rh} is
considered as alternative regularisation procedure.

Concerning non-perturbative corrections of the type $\Lambda^p/M^p$,
where $\Lambda$ is the non-perturbative scale of QCD and $M$ is the
order of magnitude of the momentum transfer in the process, we note
that the dominant power corrections are linear, for instance, in the
case of hadronic final states of electron-positron annihilation,
whereas they are expected to be quadratic for the Drell-Yan \pt{}
distribution at large
$p_T$~\cite{FerrarioRavasio:2020guj,Caola:2021kzt}, or, equivalently,
in the limit of small $b$~\cite{Tafat:2001in}.
In the small $\pt$ region, the non-perturbative corrections are
expected to become linear below some
scale~\cite{Vladimirov:2020umg,Collins:2014jpa}, which is estimated of
the order
$\mathcal{O}$(1~GeV) in Ref.~\cite{Schweitzer:2012hh}. Determinations
of non-perturbative TMD functions from fits to Drell-Yan and
semi-inclusive deep inelastic scattering (SIDIS) data further confirm a transition from quadratic to
linear behaviour below a scale which is of order $\mathcal{O}$(1.5~GeV) for
$Z$-boson production at the Tevatron~\cite{Scimemi:2019cmh}.
The $Z$-boson \pt{} distribution has negligible sensitivity to
non-perturbative corrections below such a small scale. Accordingly,
non-perturbative QCD effects are included in this analysis in the
form of a Gaussian form factor $S_{\textrm{NP}}=\exp\{-g\,b^2\}$,
which corresponds to a quadratic ansatz for the non-perturbative
corrections.

At N$^3$LL+$\mathcal{O}(\as^3)$ accuracy in the small-\qt{} region
(i.e.\ including all the $\mathcal{O}(\as^3)$ terms) we have
included in the calculation the functions $g^{(4)}$ and ${\cal H}_V^{(3)}$ in Eqs.~(\ref{eq:hexpan}) and~(\ref{eq:exponent}).
The asymptotic term $\textrm{d}\sigma^{\textrm{asy}}$ and the \Vboson+jet finite-order
cross section $\textrm{d}\sigma^{\textrm{f.o.}}$ are evaluated at
$\mathcal{O}(\as^3)$. The $\mathcal{O}(\as^3)$ term of the \Vboson+jet
cross section predictions was computed with
MCFM~\cite{Boughezal:2015ded,Neumann:2022lft}, using a lower cutoff of
$\pt = 5$~GeV, and the corresponding
$\textrm{d}\sigma^{\textrm{f.o.}}-\textrm{d}\sigma^{\textrm{asy}}$
matching correction, which is as large as $-1\%$ in the Sudakov region,
was extrapolated down to $\pt = 0$ by interpolating the corrections
with their expected quadratic dependence on
$\pt/m_Z$~\cite{Camarda:2021jsw}, i.e. with the function
$(\pt/m_Z)^2\sum_i c_i \log^i(\pt/m_Z)$ including a set of free parameters
 $c_i$ (see also Refs.~\cite{Billis:2021ecs,Chen:2022cgv} for
similar parametrisations).

The running of the strong coupling is evaluated at four loops~\cite{vanRitbergen:1997va,Czakon:2004bu}
consistently in all parts of the calculation. The PDFs are
interpolated with LHAPDF~\cite{Buckley:2014ana} at the factorisation
scale $\mu_F$, and evolved backward using the next-to-next-to-leading order
(NNLO) solution of the evolution equation, as implemented in
Ref.~\cite{Vogt:2004ns},
and four-loops running of the strong coupling. As shown
in Appendix A of Ref.~\cite{Bozzi:2005wk}, such a procedure consistently resums the
N${^3}$LL contributions to the form factor.
The number of active flavours is set to five in all the coefficients
entering the calculation, and in the evolution of the PDFs.
In order to assess the impact of charm and bottom thresholds in the PDF
evolution, an alternative forward PDF evolution with variable-flavour
number scheme is used, and the difference with respect to the nominal
five-flavour backward evolution is considered as an uncertainty.
The predicted cross sections depend on three unphysical scales: the
renormalisation scale $\mu_R$, the factorisation scale $\mu_F$, and the resummation
scale $Q$, which parametrises the arbitrariness in the resummation
procedure. The central value of the scales is set to the invariant mass
of the lepton pair $m_{\ell\ell}$. We note that within the
transverse-momentum resummation formalism of
Refs~\cite{Bozzi:2005wk,Bozzi:2010xn,Catani:2015vma} the $\mu_R$,
$\mu_F$, and $Q$ scales have to be set of the order of the hard scale
of the process $m_{\ell\ell}$ and do not depend on the transverse
momentum of the $Z$ boson.
The electroweak parameters are set according to the $G_\mu$ scheme,
in which the Fermi coupling constant $G_\textrm{F}$, the $W$-boson mass $m_W$,
and the $Z$-boson mass $m_Z$ are set to the input values
$G_\textrm{F} = 1.1663787 \cdot 10^{-5}$~GeV$^{-2}$, $m_W =
80.385$~GeV, $m_Z =
91.1876$~GeV~\cite{ParticleDataGroup:2020ssz}, whereas the weak-mixing
angle and the QED coupling are calculated at tree level.

The statistical analysis for the determination of $\asmz$ is
performed with the xFitter framework~\cite{Alekhin:2014irh}. The dependence of PDFs on the value of
$\asmz$ is accounted for by using corresponding
$\as$-series of PDF sets.
The value of $\asmz$ is determined by minimising a $\chi^2$
function which includes both the experimental uncertainties and the
theoretical uncertainties arising from PDF variations:
\begin{eqnarray}
\nonumber \lefteqn{\chi^2(\beta_{\textrm{exp}},\beta_{\textrm{th}}) = }  \\ 
&& \nonumber \sum_{i=1}^{N_{\textrm{data}}} \frac{\textstyle \left( \sigma^{\textrm{exp}}_i + \sum_j \Gamma^{\textrm{exp}}_{ij} \beta_{j,\textrm{exp}} - \sigma^{\textrm{th}}_i - \sum_k \Gamma^{\textrm{th}}_{ik}\beta_{k,\textrm{th}} \right)^2}{\Delta_i^2} \\
&&   + \sum_j \beta_{j,\textrm{exp}}^2 + \sum_k \beta_{k,\textrm{th}}^2\,.   \label{eq:chi2prof}
\end{eqnarray}
The correlated experimental and theoretical uncertainties are included
using the nuisance parameter vectors $\beta_{\textrm{exp}}$
and $\beta_{\textrm{th}}$, respectively. Their influence on
the data and theory predictions is described by the $\Gamma^{\textrm{exp}}_{ij}$ and $\Gamma^{\textrm{th}}_{ik}$ matrices. The index $i$ runs
over all $N_{\textrm{data}}$ data points, whereas the index $j$ ($k$)
corresponds to the experimental (theoretical) uncertainty nuisance
parameters. The measurements and the uncorrelated experimental
uncertainties are given by $\sigma^{\textrm{exp}}_i$ and $\Delta_i$\,, respectively, and
the theory predictions are $\sigma_i^{\textrm{th}}$.
At each value of $\asmz$, the PDF uncertainties
are Hessian profiled according to
Eq.~(\ref{eq:chi2prof})~\cite{HERAFitterdevelopersTeam:2015cre}.
The parameter $g$ of the Gaussian non-perturbative form factor is left free in the
fit by adding $g$ variations in Eq.~(\ref{eq:chi2prof}) as an
unconstrained nuisance parameter. The region of $Z$-boson transverse
momentum $\pt < 30$~GeV is considered in the fit. Initial-state
radiation of photons (QED ISR) is estimated at leading-logarithmic accuracy with \textsc{Pythia8}~\cite{Sjostrand:2014zea} and the AZ tune
of parton shower parameters~\cite{ATLAS:2014alx}, and the predictions are corrected with a bin-by-bin
multiplicative factor. The effect on $\asmz$ of including
these corrections is $\delta\asmz = -0.0006$. Uncertainties are
estimated with initial-state photon radiation at next-to-leading
logarithmic accuracy~\cite{Cieri:2018sfk}.

\section*{Results}
\begin{figure}
\includegraphics[width=0.45\textwidth]{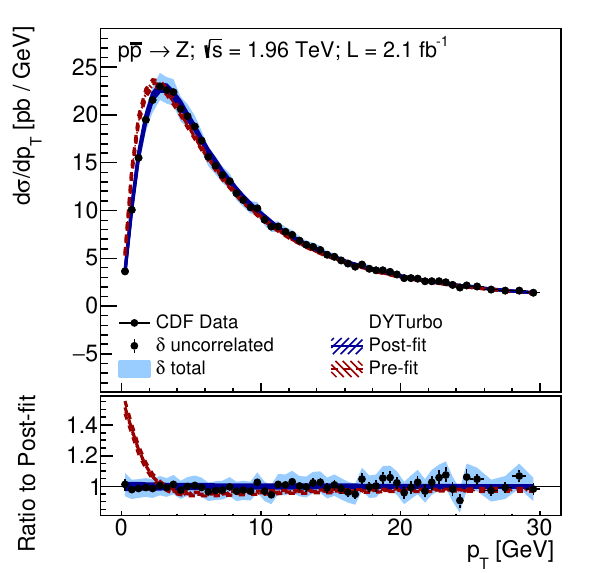}
\caption{Comparison of N$^3$LL+$\mathcal{O}(\as^3)$ \texttt{DYTurbo} predictions to the measured
$Z$-boson transverse-momentum distribution. The settings of the
  pre- and post-fit predictions are $\asmz = 0.118$, $g =
  0$~GeV$^{2}$, and $\asmz = 0.1190$, $g = 0.66$~GeV$^{2}$,
  respectively. The dashed bands represent the PDF uncertainty of the
  NNPDF4.0 PDF set.\label{fig:postfit}}
\end{figure}

\begin{figure}
\includegraphics[width=0.45\textwidth]{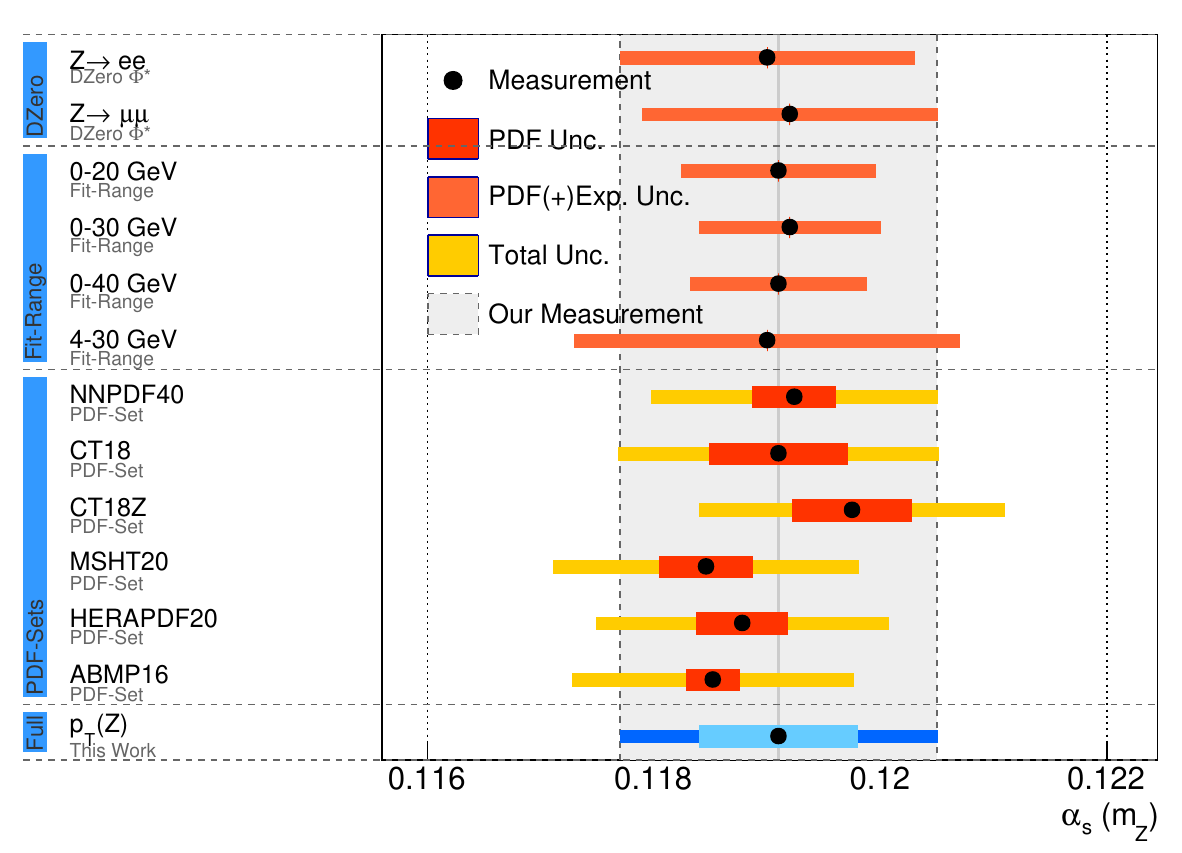}
\caption{Comparison of the $\asmz$ determination from the
  $Z$-boson transverse-momentum distribution with varying fit range, with various different
  PDF sets, and with measurements performed with the D0 detector.\label{fig:PDFs}}
\end{figure}

The determination of $\asmz$ with the Hessian conversion~\cite{Carrazza:2015aoa} of the NNLO
PDF set NNPDF4.0~\cite{NNPDF:2021njg}
yields $\asmz = 0.1192$, with a statistical uncertainty of
$\pm 0.0007$, a systematic experimental uncertainty
of $\pm 0.0001$, and a PDF uncertainty of
$\pm 0.0004$. The value of $g$ determined in the fit
is $g = 0.66 \pm 0.05$~GeV$^2$, with a correlation to \asmz{} of
$-0.8$. When performing a fit with fixed value of $g$, the
uncertainties on \asmz{} are reduced by $30\%$, yielding an estimate
for the uncertainty contribution from non-perturbative QCD effects of
$\pm 0.0006$.
The value of the $\chi^2$ function
at minimum is 41 per 53 degrees of freedom.
The pre- and post-fit predictions are compared to the measured
$Z$-boson transverse-momentum distribution in Fig.~\ref{fig:postfit}.
Various alternative NNLO
PDF sets are considered: CT18~\cite{Hou:2019efy}, CT18Z, MSHT20~\cite{Bailey:2020ooq},
HERAPDF2.0~\cite{H1:2015ubc}, and ABMP16~\cite{Alekhin:2017kpj}. The determined values of
$\asmz$ range from a minimum of $0.1185$ with the MSHT20 PDF
set to a maximum of $0.1198$ with the CT18Z PDF set. The midpoint value
in this range of $\asmz = 0.1191$ is considered as nominal
result, and the PDF envelope of $\pm 0.0007$ as an additional source
of uncertainty.
The determination of \asmz{} from the various different NNLO PDF sets is
shown in Fig.~\ref{fig:PDFs}. The approximate N$^3$LO MSHT20 PDF
set~\cite{McGowan:2022nag} is also considered, using predictions at approximate
next-to-next-to-next-to-next-to-leading-logarithmic (N$^4$LL)
accuracy~\cite{Camarda:2023dqn}, yielding a value of $\asmz = 0.1184$.
Determinations of $\asmz$ at hadron colliders are exposed to possible
biases unless the PDFs are determined simultaneously along with
$\asmz$~\cite{Forte:2020pyp}. Nonetheless, $\asmz$ determinations from
single or limited hadron collider datasets based on existing PDF sets,
are interesting to study in detail the sensitivity to \asmz{} of a particular
observable and the associated theoretical uncertainties.
The Hessian profiling employed in this analysis provides an
approximation to a PDF determination which relies on the accuracy of
the quadratic approximation around the minimum~\cite{Paukkunen:2014zia} (see Appendix~\ref{app:pdffit} for details).
In all the cases considered in this analysis, pulls and constraints of
the nuisance parameters associated to the PDF uncertainties are below
$20\%$ and $10\%$, respectively, indicating that the new minimum of
the profiled PDFs is very close to the original minimum, which gives
confidence in the validity of the quadratic approximation.

With the aim of further testing the validity of the Hessian profiling
approximation, a simultaneous fit of PDFs, \asmz, and the non-perturbative
parameter $g$ is performed. The combined neutral and charged current
deep inelastic scattering (DIS) cross-section data from the H1 and
ZEUS experiments at the HERA collider~\cite{H1:2015ubc} are included
in the fit, with a minimum squared four-momentum transfer $Q^2$ of
3.5~GeV$^2$, together with the $Z$-boson trans\-verse-momentum
distribution measured by CDF.
The light-quark coefficient functions of the DIS cross sections are
calculated in the $\overline{MS}$ scheme~\cite{Weinberg:1973xwm}, and
with the renormalisation and factorisation scales set to the squared four-momentum
transfer $Q^2$.
The heavy quarks $c$ and $b$ are dynamically generated, and the
corresponding coefficient functions for the neutral-current processes
with $\gamma^*$ exchange are calculated in the general-mass
variable-flavour-number (VFN)
scheme~\cite{Thorne:1997ga,Thorne:2006qt,Thorne:2012az}, with up to
five active quark flavours.
The charm mass is set to $m_c = 1.43$~GeV, and the bottom mass to $m_b
= 4.50$~GeV~\cite{H1:2015ubc}. For the charged-current processes the
heavy quarks are treated as massless.
The PDFs for the gluon, $u$-valence, $d$-valence, $\bar{u}$, $\bar{d}$
quark densities are parameterised at the input scale
$Q^2_0=1.9$~GeV$^2$ with the parametrisation of Ref.~\cite{H1:2015ubc}.
The contribution of the $s$-quark density is taken to be proportional to the
$\bar{d}$-quark density by setting $x\bar{s}(x) = r_s x\bar{d}(x)$,
with $r_s=0.67$.
The determined value of \asmz{} from
this fit is $0.1184 \pm 0.0006$, where the quoted uncertainty is the
uncertainty from the fit, which includes experimental and PDF uncertainties.
The value of \asmz{} is in agreement with the determinations based on
the Hessian profiling approach.

The alternative fits with different PDF sets and the simultaneous fit
of PDFs and \asmz{} are summarised in Table~\ref{tab:pdfs}.
\begin{table*}
  \begin{center}
    \caption{\label{tab:pdfs} Alternative fits of \asmz{} with different PDF sets.}
    \begin{tabular}{ccccccc}
      \toprule
                                                  & \asmz & $g$ [GeV$^2$] & $\chi^2$/dof \\
      \midrule
      NNPDF4.0                                    & $0.1192 \pm 0.0008$ & $0.66 \pm 0.05$        & 41/53 \\
      CT18                                        & $0.1189 \pm 0.0010$ & $0.67 \pm 0.05$        & 40/53 \\
      CT18Z                                       & $0.1198 \pm 0.0009$ & $0.62 \pm 0.05$        & 41/53 \\
      MSHT20                                      & $0.1185 \pm 0.0009$ & $0.72 \pm 0.05$        & 40/53 \\
      HERAPDF2.0                                  & $0.1188 \pm 0.0008$ & $0.69 \pm 0.05$        & 40/53 \\
      ABMP16                                      & $0.1185 \pm 0.0007$ & $0.62 \pm 0.05$        & 42/53 \\
      \midrule
      MSHT20an3lo (N$^4$LL)                       & $0.1184 \pm 0.0009$ & $0.73 \pm 0.05$        & 40/53 \\
      \midrule
      PDF fit                                     & $0.1184 \pm 0.0006$ & $0.71 \pm 0.05$        & 1405/1184 \\
      \bottomrule
    \end{tabular}
  \end{center}
\end{table*}

Missing higher order uncertainties are estimated
through in\-de\-pen\-dent variations of $\mu_R$, $\mu_F$ and $Q$ in the
range $m_{\ell\ell}/2 \leq \{ \mu_R, \mu_F, Q \} \leq 2m_{\ell\ell}$
with the constraints $0.5\leq \{ \mu_F/\mu_R, Q/\mu_R, Q/\mu_F \}\leq
2$, leading to 14 variations.
The determined values of $\asmz$ range from a minimum of
$0.1183$ to a maximum of $0.1196$ with respect to the value at the
central scale choice of $\asmz = 0.1192$, yielding a scale-variation envelope of
$^{+0.0004}_{-0.0009}$.
The alternative fits with different choices of the QCD scales are summarised in Table~\ref{tab:scales}.

\begin{table*}
  \begin{center}
    \caption{\label{tab:scales} Alternative fits of \asmz{} with different choices of the renormalisation ($\mu_R$), factorisation ($\mu_F$) and resummation ($Q$) scales.}
    \begin{tabular}{ccccccc}
      \toprule
$\mu_R/m_{\ell\ell}$ & $\mu_F/m_{\ell\ell}$ & $Q/m_{\ell\ell}$ & \asmz & $g$ [GeV$^2$] & $\chi^2$/dof \\
\midrule
      1       &     1      &   1       & $0.1192 \pm 0.0008$     & $0.66 \pm 0.05$    & 41/53 \\
      1       &     1      &   2       & $0.1183 \pm 0.0007$     & $0.77 \pm 0.05$    & 40/53 \\
      1       &     1      &   0.5     & $0.1196 \pm 0.0008$     & $0.57 \pm 0.05$    & 42/53 \\
      1       &     2      &   1       & $0.1194 \pm 0.0008$     & $0.66 \pm 0.05$    & 41/53 \\
      1       &     2      &   2       & $0.1183 \pm 0.0007$     & $0.77 \pm 0.05$    & 41/53 \\
      1       &     0.5    &   1       & $0.1193 \pm 0.0008$     & $0.68 \pm 0.05$    & 42/53 \\
      1       &     0.5    &   0.5     & $0.1196 \pm 0.0008$     & $0.59 \pm 0.05$    & 42/53 \\
      2       &     1      &   1       & $0.1193 \pm 0.0008$     & $0.67 \pm 0.05$    & 42/53 \\
      2       &     1      &   2       & $0.1194 \pm 0.0008$     & $0.70 \pm 0.05$    & 41/53 \\
      2       &     2      &   1       & $0.1192 \pm 0.0008$     & $0.65 \pm 0.05$    & 42/53 \\
      2       &     2      &   2       & $0.1192 \pm 0.0008$     & $0.67 \pm 0.05$    & 41/53 \\
      0.5     &     1      &   1       & $0.1184 \pm 0.0007$     & $0.75 \pm 0.05$    & 42/53 \\
      0.5     &     1      &   0.5     & $0.1192 \pm 0.0007$     & $0.64 \pm 0.05$    & 41/53 \\
      0.5     &     0.5    &   1       & $0.1183 \pm 0.0007$     & $0.75 \pm 0.05$    & 42/53 \\
      0.5     &     0.5    &   0.5     & $0.1192 \pm 0.0007$     & $0.64 \pm 0.05$    & 42/53 \\
      \bottomrule
    \end{tabular}
  \end{center}
\end{table*}

Fits without the $\mathcal{O}(\as^3)$
matching corrections yield a central value which is 0.0005 lower, and
an increase in the half envelope of scale variations from 0.0007 to
0.0009, which is consistent with the observed shift.
Systematic uncertainties in the $\mathcal{O}(\as^3)$
matching corrections are estimated by raising the lower cutoff from
$\pt = 5$~GeV to $\pt = 10$~GeV. The difference of $0.0001$ with
respect to the nominal fit is considered as a source of uncertainty.
Statistical uncertainties in the $\mathcal{O}(\as^3)$
matching corrections are estimated with a set of 1000 replicas of the
matching corrections generated by fluctuating them within their
numerical uncertainties. The upper and lower limits of the 68$\%$
confidence level envelope of interpolations to the replicas are used for the estimate of
the statistical uncertainty, yielding less than $\pm 0.0001$. Further details are provided in Appendix~\ref{app:match}.

Uncertainties in the modelling of the non-perturbative form factor are
estimated by performing four alternative fits with: a value of $b_{\textrm{lim}} = 2$~GeV$^{-1}$ in
the $b_*$ regularisation procedure; the minimal prescription,
which corresponds to the limit $b_{\textrm{lim}} \to \infty$; using an
additional quartic term $\exp\{-q\,b^4\}$ with $q = 0.1$~GeV$^4$;
using the additional term $\exp\{-g_{k}\}$ with
$g_k = g_0\big(1-\exp\big[-\frac{\textrm{C}_{\textrm{F}}}{\pi g_0
    b_{\textrm{lim}}^2}\big]\big)\log(m_{\ell\ell}^2/Q_0^2)$
with $g_0 = 0.3$, $Q_0 = 1$~GeV, and $b_{\textrm{lim}} = 2$~GeV$^{-1}$~\cite{Collins:2014jpa}, where $C_F$ is the colour-factor associated with gluon emission from a quark.
The alternative fits yield variations of \asmz{} in the range of
$\pm 0.0007$, which is considered as an uncertainty. In
the alternative fits, the parameter of the Gaussian non-perturbative form factor
ranges from $g = 0.42$~GeV$^{2}$ in the case of the minimal
prescription to $g = 0.83$~GeV$^{2}$ in the case of the fit with
$b_{\textrm{lim}} = 2$~GeV$^{-1}$, in agreement with values obtained
by global fits~\cite{Konychev:2005iy,Bacchetta:2019sam,Sun:2014dqm}, and corresponding to values of
average primordial $k_T^2$ of the partons, $\langle k_T^2 \rangle = 2g$~\cite{Parisi:1979se,Hirai:2012ada}, in the range $0.8$ to $1.7$
GeV$^2$. Such values are generally large for non-perturbative effects
within a bound state with a mass of $1$~GeV as the proton. However the fitted
values of $g$ also accounts for power corrections related to the
regularisation procedure of the perturbative form factor, to the
perturbative evolution of the non-perturbative form factor from low scales to $m_Z$,
and to yet uncalculated higher-order corrections.
A fit in which the NNPDF4.0 PDF set is evolved
with a variable-flavour number scheme yields $\delta\asmz =
-0.0003$, which is considered as an additional source of uncertainty.
The alternative fits with different non-perturbative and heavy flavour models are summarised in Table~\ref{tab:np}.
\begin{table*}
  \begin{center}
    \caption{\label{tab:np} Alternative fits of \asmz{} with different non-perturbative and heavy flavour models.}
    \begin{tabular}{ccccccc}
      \toprule
                                                  & \asmz & $g$ [GeV$^2$] & $\chi^2$/dof \\
      \midrule
      $b_{\textrm{lim}} = 2$~GeV$^{-1}$             & $0.1187 \pm 0.0007$ & $0.83 \pm 0.05$        & 43/53 \\
      $b_{\textrm{lim}} \to \infty$                & $0.1199 \pm 0.0008$ & $0.42 \pm 0.05$         & 41/53 \\
      $g_k$                                       & $0.1186 \pm 0.0008$ & $0.65 \pm 0.05$        & 46/53 \\
      $q = 0.1$~GeV$^4$                           & $0.1197 \pm 0.0008$ & $0.51 \pm 0.05$        & 41/53 \\
      \midrule
      VFN PDF evolution                           & $0.1190 \pm 0.0007$ & $0.71 \pm 0.05$        & 59/53 \\
      \bottomrule
    \end{tabular}
  \end{center}
\end{table*}

A fit with NLL initial-state radiation of photons yields a difference
on \asmz with respect to the \textsc{Pythia8} modelling of less than
$0.0001$, which is considered as an additional source of uncertainty.

The stability of the results upon variations of the fit range is
tested by performing fits in the regions of $Z$-boson transverse
momentum $\pt < 20$~GeV and $\pt < 40$~GeV. The spread in the
determined values of $\asmz$ is at the level of
$\pm 0.0001$ and the uncertainty of the fit increases from $\pm0.0007$
to $\pm0.0008$. Since the region $20 < \pt < 40$~GeV is sensitive to the
matching of the resummed cross section to the fixed order prediction,
this test provides a confirmation that the result is largely
independent from the matching corrections in this region.
Uncertainties associated to the stability of the fit results with
respect to variations of the upper limit of the fit range are considered negligible.
The fit range is also varied by excluding the low
transverse-momentum region. The range is reduced up to $4 < \pt <
30$~GeV, with a spread in the values of $\asmz$ at the level of
$\pm 0.0002$, and an increase in the uncertainty of the fit from $\pm0.0008$
to $\pm0.0016$. For the fit in the range $4 < \pt < 30$~GeV the value
of $g$ is determined as $0.3 \pm 0.3$~GeV$^2$ and the correlation between \asmz{} and
$g$ is reduced from $-0.8$ to $-0.4$. Since the low transverse-momentum region is the most
sensitive to the non-perturbative QCD effects, this test provides a
validation of the model for the non-perturbative form factor.
The spread of $\pm 0.0002$ from variations of the lower limit of the
fit range is considered as an additional source of uncertainty.

A consistency check of the \asmz{} determination was performed using
cross sections measured with the D0 detector~\cite{D0:2010qhp}. The fit to the
D0 data in the $Z$-boson rapidity range $|y| < 1$ yields value of
$\asmz = 0.1190 \pm 0.0013$ in the electron decay channel and $\asmz =
0.1192 \pm 0.0013$ in the muon decay channel, where the quoted
uncertainties include experimental and PDF uncertainties. The D0
measurement, which was performed on the variable $\phi^{*}_{\eta}$, is
extrapolated to the transverse-momentum \pt. The extrapolation
procedure has associated uncertainties
which were not estimated in the analysis. The determined values of \asmz{}
are compatible with the CDF result within experimental
uncertainties. Determinations of \asmz{} with varying fit range and
with cross sections measured with the D0 detector are shown in Fig.~\ref{fig:PDFs}.

A summary of the uncertainties in the determination of \asmz{} is shown
in Table~\ref{tab:syst}.
\begin{table*}
  \begin{center}
    \caption{\label{tab:syst} Summary of the uncertainties for the
    determination of \asmz, in units of $10^{-3}$.}
    \begin{tabular}{ccccccc}
      \toprule
      Statistical uncertainty                     & \multicolumn{2}{c}{$\pm0.7$}\\
      Experimental systematic uncertainty         & \multicolumn{2}{c}{$\pm0.1$}\\
      PDF uncertainty (NNPDF4.0)                  & \multicolumn{2}{c}{$\pm0.4$}\\
      PDF uncertainty (envelope of PDFs)          & \multicolumn{2}{c}{$\pm0.7$}\\
      Scale variations  uncertainties             & +0.4  & -0.9 \\
      Matching at $\mathcal{O}(\as^3)$            & \multicolumn{2}{c}{$\pm0.1$}\\
      Non-perturbative model                      & \multicolumn{2}{c}{$\pm0.7$}\\
      Flavour model                               & 0 & -0.3\\
      QED ISR                                     & \multicolumn{2}{c}{$<\pm0.1$}\\
      Lower limit of fit range                    & \multicolumn{2}{c}{$\pm0.2$}\\
      \midrule                                    
      Total                                       & +1.3 & -1.6\\
      \bottomrule
    \end{tabular}
  \end{center}
\end{table*}

\section*{Conclusions}
In summary, the value of the strong-coupling constant determined in this
analysis is $\asmz = 0.1191^{+0.0013}_{-0.0016}$, with a statistical
uncertainty of $\pm 0.0007$, an experimental systematic uncertainty of
$\pm 0.0001$, a PDF uncertainty of $\pm 0.0008$, missing higher order
uncertainties of $^{+0.0004}_{-0.0009}$, and additional theory
uncertainties (non-perturbative model, flavour scheme, matching
corrections, photon initial-state radiation) of
$\pm{0.0008}$. The strong-coupling constant is
also determined in a simultaneous PDF-fit determination including DIS
cross-sec\-tion data from the H1 and ZEUS experiments at the HERA
collider. When considering the fit uncertainties of $\pm 0.0006$ and
all the other relevant uncertainties listed in Table~\ref{tab:syst},
the result of this determination is $\asmz = 0.1184 ^{+0.0013}_{-0.0015}$.

\begin{figure}
\includegraphics[width=0.45\textwidth]{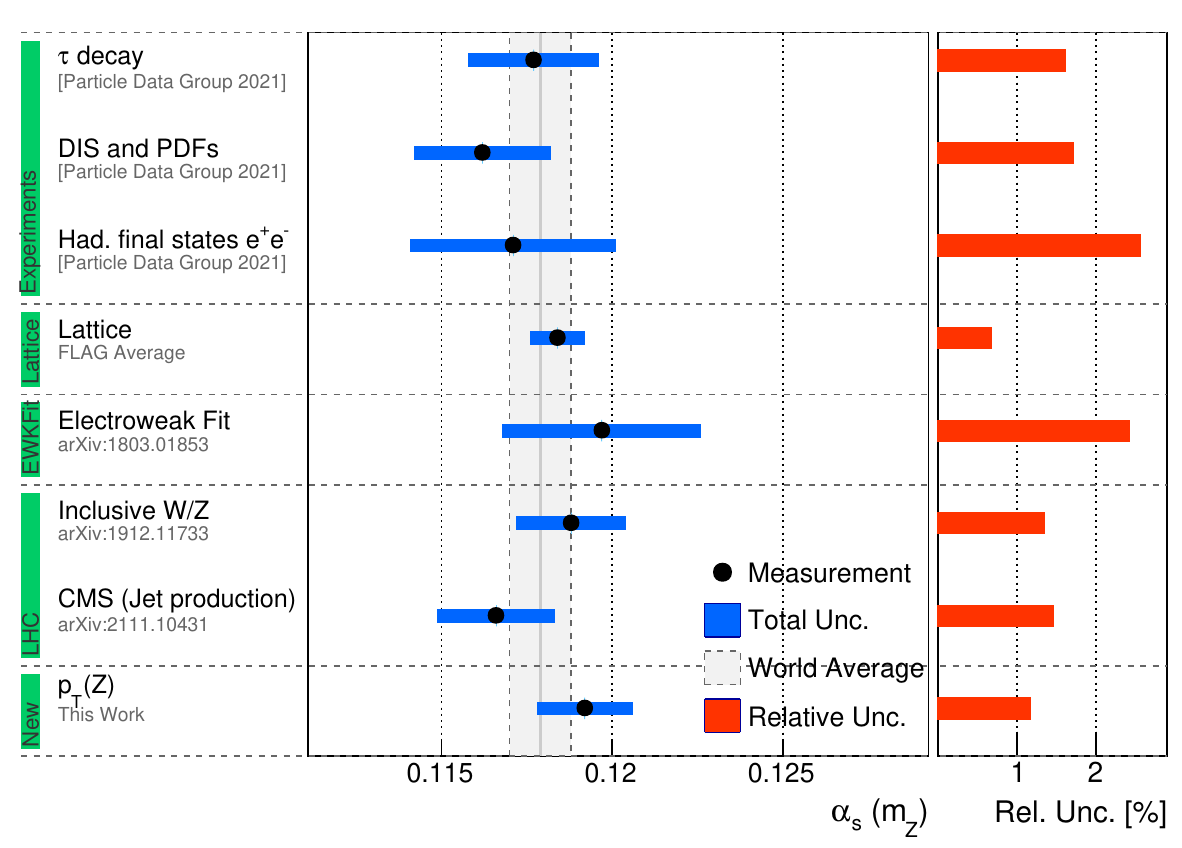}
\caption{Comparison of the $\asmz$ determination from the
  $Z$-boson transverse-momentum distribution to other determinations
  and to the world-average value.\label{fig:overview}}
\end{figure}

We have performed a determination of $\asmz$ from the
$Z$-boson transverse-momentum distribution measured at the Tevatron
collider, in the low-momentum region of $\pt < 30$~GeV. This analysis
represents the first determination using QCD resummed theory predictions based on
a semi-inclusive observable at hadron-hadron colliders\footnote{
Analogous QCD resummed theory predictions in electron-positron
collisions were used to determine \asmz{} at LEP~\cite{Bethke:1992jn,Catani:1992ua,Abbate:2010xh,Gehrmann:2012sc,Hoang:2015hka,Verbytskyi:2019zhh}.}.
The PDF uncertainties are estimated with a conservative
approach, including the envelope of six different PDF sets, and with a
Hessian profiling procedure, which avoids possible biases in the
treatment of PDF uncertainties. Missing
higher order uncertainties are estimated with the standard approach of
computing an envelope of scale variations.
The measured value of $\asmz$ has a relative uncertainty of
$1.2\%$, and is compatible with other determinations and with the
world-average value, as illustrated in Fig.~\ref{fig:overview}.

Among hadron colliders determination, this is the most precise to
date and the first based on N$^3$LL+$\mathcal{O}(\as^3)$ predictions in perturbative QCD.

\section*{Acknowledgements}
We gratefully acknowledge Willis Sakumoto for providing the correlation
matrix of the CDF $Z$-boson transverse-momentum measurement, Tobias
Neumann for providing the MCFM NNLO $Z$+jet predictions, Pier
Monni for suggesting to include variations of the lower edge of the
fit range, Dave Soper for fruitful discussions,
and Sasha Glazov and Stefano Catani for useful comments on the
manuscript.
This research used resources of the National Energy Research Scientific
Computing Center (NERSC), a U.S. Department of Energy Office of Science
User Facility located at Lawrence Berkeley National Laboratory, operated
under Contract No. DE-AC02-05CH11231 using NERSC award HEP-ERCAP0021890.

\bibliographystyle{utphys} 
\bibliography{zptcdf}

\providecommand{\href}[2]{#2}\begingroup\raggedright\begin{thebibliography}{10}

\bibitem{ParticleDataGroup:2020ssz}
{\bfseries Particle Data Group} Collaboration, P.~A. Zyla {\em et~al.},
  ``{Review of Particle Physics},''
  \href{http://dx.doi.org/10.1093/ptep/ptaa104}{{\em PTEP} {\bfseries 2020}
  no.~8, (2020) 083C01}.

\bibitem{Salam:2017qdl}
G.~P. Salam, {\em {The strong coupling: a theoretical perspective}},
  \href{http://dx.doi.org/10.1142/9789813238053_0007}{pp.~101--121}.
\newblock 2019.
\newblock \href{http://arxiv.org/abs/1712.05165}{{\ttfamily arXiv:1712.05165
  [hep-ph]}}.

\bibitem{Aoki:2021kgd}
{\bfseries Flavour Lattice Averaging Group (FLAG)} Collaboration, Y.~Aoki {\em
  et~al.}, ``{FLAG Review 2021},''
  \href{http://dx.doi.org/10.1140/epjc/s10052-022-10536-1}{{\em Eur. Phys. J.
  C} {\bfseries 82} no.~10, (2022) 869},
  \href{http://arxiv.org/abs/2111.09849}{{\ttfamily arXiv:2111.09849
  [hep-lat]}}.

\bibitem{Blumlein:2006be}
J.~Blumlein, H.~Bottcher, and A.~Guffanti, ``{Non-singlet QCD analysis of deep
  inelastic world data at O(alpha(s)**3)},''
  \href{http://dx.doi.org/10.1016/j.nuclphysb.2007.03.035}{{\em Nucl. Phys. B}
  {\bfseries 774} (2007) 182--207},
  \href{http://arxiv.org/abs/hep-ph/0607200}{{\ttfamily arXiv:hep-ph/0607200}}.

\bibitem{Jimenez-Delgado:2014twa}
P.~Jimenez-Delgado and E.~Reya, ``{Delineating parton distributions and the
  strong coupling},'' \href{http://dx.doi.org/10.1103/PhysRevD.89.074049}{{\em
  Phys. Rev. D} {\bfseries 89} no.~7, (2014) 074049},
  \href{http://arxiv.org/abs/1403.1852}{{\ttfamily arXiv:1403.1852 [hep-ph]}}.

\bibitem{Alekhin:2017kpj}
S.~Alekhin, J.~Bl\"umlein, S.~Moch, and R.~Placakyte, ``{Parton distribution
  functions, $\alpha_s$, and heavy-quark masses for LHC Run II},''
  \href{http://dx.doi.org/10.1103/PhysRevD.96.014011}{{\em Phys. Rev. D}
  {\bfseries 96} no.~1, (2017) 014011},
  \href{http://arxiv.org/abs/1701.05838}{{\ttfamily arXiv:1701.05838
  [hep-ph]}}.

\bibitem{Abbate:2010xh}
R.~Abbate, M.~Fickinger, A.~H. Hoang, V.~Mateu, and I.~W. Stewart, ``{Thrust at
  $N^{3}LL$ with Power Corrections and a Precision Global Fit for
  $\alpha_{s}(mZ)$},'' \href{http://dx.doi.org/10.1103/PhysRevD.83.074021}{{\em
  Phys. Rev. D} {\bfseries 83} (2011) 074021},
  \href{http://arxiv.org/abs/1006.3080}{{\ttfamily arXiv:1006.3080 [hep-ph]}}.

\bibitem{Gehrmann:2012sc}
T.~Gehrmann, G.~Luisoni, and P.~F. Monni, ``{Power corrections in the
  dispersive model for a determination of the strong coupling constant from the
  thrust distribution},''
  \href{http://dx.doi.org/10.1140/epjc/s10052-012-2265-x}{{\em Eur. Phys. J. C}
  {\bfseries 73} no.~1, (2013) 2265},
  \href{http://arxiv.org/abs/1210.6945}{{\ttfamily arXiv:1210.6945 [hep-ph]}}.

\bibitem{Hoang:2015hka}
A.~H. Hoang, D.~W. Kolodrubetz, V.~Mateu, and I.~W. Stewart, ``{Precise
  determination of $\alpha_s$ from the $C$-parameter distribution},''
  \href{http://dx.doi.org/10.1103/PhysRevD.91.094018}{{\em Phys. Rev. D}
  {\bfseries 91} no.~9, (2015) 094018},
  \href{http://arxiv.org/abs/1501.04111}{{\ttfamily arXiv:1501.04111
  [hep-ph]}}.

\bibitem{Baikov:2008jh}
P.~A. Baikov, K.~G. Chetyrkin, and J.~H. Kuhn, ``{Order alpha**4(s) QCD
  Corrections to Z and tau Decays},''
  \href{http://dx.doi.org/10.1103/PhysRevLett.101.012002}{{\em Phys. Rev.
  Lett.} {\bfseries 101} (2008) 012002},
  \href{http://arxiv.org/abs/0801.1821}{{\ttfamily arXiv:0801.1821 [hep-ph]}}.

\bibitem{Mateu:2017hlz}
V.~Mateu and P.~G. Ortega, ``{Bottom and Charm Mass determinations from global
  fits to $Q\bar{Q}$ bound states at N$^3$LO},''
  \href{http://dx.doi.org/10.1007/JHEP01(2018)122}{{\em JHEP} {\bfseries 01}
  (2018) 122}, \href{http://arxiv.org/abs/1711.05755}{{\ttfamily
  arXiv:1711.05755 [hep-ph]}}.

\bibitem{Peset:2018ria}
C.~Peset, A.~Pineda, and J.~Segovia, ``{The charm/bottom quark mass from heavy
  quarkonium at N$^{3}$LO},''
  \href{http://dx.doi.org/10.1007/JHEP09(2018)167}{{\em JHEP} {\bfseries 09}
  (2018) 167}, \href{http://arxiv.org/abs/1806.05197}{{\ttfamily
  arXiv:1806.05197 [hep-ph]}}.

\bibitem{Haller:2018nnx}
J.~Haller, A.~Hoecker, R.~Kogler, K.~M\"onig, T.~Peiffer, and J.~Stelzer,
  ``{Update of the global electroweak fit and constraints on two-Higgs-doublet
  models},'' \href{http://dx.doi.org/10.1140/epjc/s10052-018-6131-3}{{\em Eur.
  Phys. J. C} {\bfseries 78} no.~8, (2018) 675},
  \href{http://arxiv.org/abs/1803.01853}{{\ttfamily arXiv:1803.01853
  [hep-ph]}}.

\bibitem{dEnterria:2020cpv}
D.~d'Enterria and V.~Jacobsen, ``{Improved strong coupling determinations from
  hadronic decays of electroweak bosons at N$^3$LO accuracy},''
  \href{http://arxiv.org/abs/2005.04545}{{\ttfamily arXiv:2005.04545
  [hep-ph]}}.

\bibitem{CMS:2021yzl}
{\bfseries CMS} Collaboration, A.~Tumasyan {\em et~al.}, ``{Measurement and QCD
  analysis of double-differential inclusive jet cross sections in proton-proton
  collisions at $ \sqrt{s} $ = 13 TeV},''
  \href{http://dx.doi.org/10.1007/JHEP02(2022)142}{{\em JHEP} {\bfseries 02}
  (2022) 142}, \href{http://arxiv.org/abs/2111.10431}{{\ttfamily
  arXiv:2111.10431 [hep-ex]}}. [Addendum: JHEP 12, 035 (2022)].

\bibitem{ATLAS:2017qir}
{\bfseries ATLAS} Collaboration, M.~Aaboud {\em et~al.}, ``{Determination of
  the strong coupling constant $\alpha _\mathrm {s}$ from transverse
  energy\textendash{}energy correlations in multijet events at $\sqrt{s} =
  8~\hbox {TeV}$ using the ATLAS detector},''
  \href{http://dx.doi.org/10.1140/epjc/s10052-017-5442-0}{{\em Eur. Phys. J. C}
  {\bfseries 77} no.~12, (2017) 872},
  \href{http://arxiv.org/abs/1707.02562}{{\ttfamily arXiv:1707.02562
  [hep-ex]}}.

\bibitem{CMS:2018fks}
{\bfseries CMS} Collaboration, A.~M. Sirunyan {\em et~al.}, ``{Measurement of
  the $\mathrm{t}\overline{\mathrm{t}}$ production cross section, the top quark
  mass, and the strong coupling constant using dilepton events in pp collisions
  at $\sqrt{s} =$ 13 TeV},''
  \href{http://dx.doi.org/10.1140/epjc/s10052-019-6863-8}{{\em Eur. Phys. J. C}
  {\bfseries 79} no.~5, (2019) 368},
  \href{http://arxiv.org/abs/1812.10505}{{\ttfamily arXiv:1812.10505
  [hep-ex]}}.

\bibitem{CMS:2014rml}
{\bfseries CMS} Collaboration, S.~Chatrchyan {\em et~al.}, ``{Determination of
  the Top-Quark Pole Mass and Strong Coupling Constant from the $t \bar{t}$
  Production Cross Section in $pp$ Collisions at $\sqrt{s}$ = 7 TeV},''
  \href{http://dx.doi.org/10.1016/j.physletb.2013.12.009}{{\em Phys. Lett. B}
  {\bfseries 728} (2014) 496--517},
  \href{http://arxiv.org/abs/1307.1907}{{\ttfamily arXiv:1307.1907 [hep-ex]}}.
  [Erratum: Phys.Lett.B 738, 526--528 (2014)].

\bibitem{Klijnsma:2017eqp}
T.~Klijnsma, S.~Bethke, G.~Dissertori, and G.~P. Salam, ``{Determination of the
  strong coupling constant $\alpha_s(m_Z)$ from measurements of the total cross
  section for top-antitop quark production},''
  \href{http://dx.doi.org/10.1140/epjc/s10052-017-5340-5}{{\em Eur. Phys. J. C}
  {\bfseries 77} no.~11, (2017) 778},
  \href{http://arxiv.org/abs/1708.07495}{{\ttfamily arXiv:1708.07495
  [hep-ph]}}.

\bibitem{dEnterria:2019aat}
D.~d'Enterria and A.~Poldaru, ``{Extraction of the strong coupling
  $\alpha_{s}$(m$_{Z}$) from a combined NNLO analysis of inclusive electroweak
  boson cross sections at hadron colliders},''
  \href{http://dx.doi.org/10.1007/JHEP06(2020)016}{{\em JHEP} {\bfseries 06}
  (2020) 016}, \href{http://arxiv.org/abs/1912.11733}{{\ttfamily
  arXiv:1912.11733 [hep-ph]}}.

\bibitem{ATLAS:2014alx}
{\bfseries ATLAS} Collaboration, G.~Aad {\em et~al.}, ``{Measurement of the
  $Z/\gamma^*$ boson transverse momentum distribution in $pp$ collisions at
  $\sqrt{s}$ = 7 TeV with the ATLAS detector},''
  \href{http://dx.doi.org/10.1007/JHEP09(2014)145}{{\em JHEP} {\bfseries 09}
  (2014) 145}, \href{http://arxiv.org/abs/1406.3660}{{\ttfamily arXiv:1406.3660
  [hep-ex]}}.

\bibitem{ATLAS:2015iiu}
{\bfseries ATLAS} Collaboration, G.~Aad {\em et~al.}, ``{Measurement of the
  transverse momentum and $\phi ^*_{\eta }$ distributions of
  Drell\textendash{}Yan lepton pairs in proton\textendash{}proton collisions at
  $\sqrt{s}=8$ TeV with the ATLAS detector},''
  \href{http://dx.doi.org/10.1140/epjc/s10052-016-4070-4}{{\em Eur. Phys. J. C}
  {\bfseries 76} no.~5, (2016) 291},
  \href{http://arxiv.org/abs/1512.02192}{{\ttfamily arXiv:1512.02192
  [hep-ex]}}.

\bibitem{CMS:2015hyl}
{\bfseries CMS} Collaboration, V.~Khachatryan {\em et~al.}, ``{Measurement of
  the Z boson differential cross section in transverse momentum and rapidity in
  proton\textendash{}proton collisions at 8 TeV},''
  \href{http://dx.doi.org/10.1016/j.physletb.2015.07.065}{{\em Phys. Lett. B}
  {\bfseries 749} (2015) 187--209},
  \href{http://arxiv.org/abs/1504.03511}{{\ttfamily arXiv:1504.03511
  [hep-ex]}}.

\bibitem{Boughezal:2017nla}
R.~Boughezal, A.~Guffanti, F.~Petriello, and M.~Ubiali, ``{The impact of the
  LHC Z-boson transverse momentum data on PDF determinations},''
  \href{http://dx.doi.org/10.1007/JHEP07(2017)130}{{\em JHEP} {\bfseries 07}
  (2017) 130}, \href{http://arxiv.org/abs/1705.00343}{{\ttfamily
  arXiv:1705.00343 [hep-ph]}}.

\bibitem{Ball:2018iqk}
{\bfseries NNPDF} Collaboration, R.~D. Ball, S.~Carrazza, L.~Del~Debbio,
  S.~Forte, Z.~Kassabov, J.~Rojo, E.~Slade, and M.~Ubiali, ``{Precision
  determination of the strong coupling constant within a global PDF
  analysis},'' \href{http://dx.doi.org/10.1140/epjc/s10052-018-5897-7}{{\em
  Eur. Phys. J. C} {\bfseries 78} no.~5, (2018) 408},
  \href{http://arxiv.org/abs/1802.03398}{{\ttfamily arXiv:1802.03398
  [hep-ph]}}.

\bibitem{Hou:2019efy}
T.-J. Hou {\em et~al.}, ``{New CTEQ global analysis of quantum chromodynamics
  with high-precision data from the LHC},''
  \href{http://dx.doi.org/10.1103/PhysRevD.103.014013}{{\em Phys. Rev. D}
  {\bfseries 103} no.~1, (2021) 014013},
  \href{http://arxiv.org/abs/1912.10053}{{\ttfamily arXiv:1912.10053
  [hep-ph]}}.

\bibitem{Cridge:2021qfd}
T.~Cridge, L.~A. Harland-Lang, A.~D. Martin, and R.~S. Thorne, ``{An
  investigation of the $\alpha _S$ and heavy quark mass dependence in the
  MSHT20 global PDF analysis},''
  \href{http://dx.doi.org/10.1140/epjc/s10052-021-09533-7}{{\em Eur. Phys. J.
  C} {\bfseries 81} no.~8, (2021) 744},
  \href{http://arxiv.org/abs/2106.10289}{{\ttfamily arXiv:2106.10289
  [hep-ph]}}.

\bibitem{Sudakov:1954sw}
V.~V. Sudakov, ``{Vertex parts at very high-energies in quantum
  electrodynamics},'' {\em Sov. Phys. JETP} {\bfseries 3} (1956) 65--71.

\bibitem{PhysRevLett.25.902.2}
S.~D. Drell and T.-M. Yan, ``Massive lepton-pair production in hadron-hadron
  collisions at high energies,''
  \href{http://dx.doi.org/10.1103/PhysRevLett.25.902.2}{{\em Phys. Rev. Lett.}
  {\bfseries 25} (Sep, 1970) 902--902}.

\bibitem{Camarda:2021ict}
S.~Camarda, L.~Cieri, and G.~Ferrera, ``{Drell\textendash{}Yan lepton-pair
  production: qT resummation at N3LL accuracy and fiducial cross sections at
  N3LO},'' \href{http://dx.doi.org/10.1103/PhysRevD.104.L111503}{{\em Phys.
  Rev. D} {\bfseries 104} no.~11, (2021) L111503},
  \href{http://arxiv.org/abs/2103.04974}{{\ttfamily arXiv:2103.04974
  [hep-ph]}}.

\bibitem{Re:2021con}
E.~Re, L.~Rottoli, and P.~Torrielli, ``{Fiducial Higgs and Drell-Yan
  distributions at N$^3$LL$^\prime$+NNLO with RadISH},''
  \href{http://arxiv.org/abs/2104.07509}{{\ttfamily arXiv:2104.07509
  [hep-ph]}}.

\bibitem{Ju:2021lah}
W.-L. Ju and M.~Sch\"onherr, ``{The q$_{T}$ and
  \ensuremath{\Delta}\ensuremath{\phi} spectra in W and Z production at the LHC
  at N$^{3}$LL'+N$^{2}$LO},''
  \href{http://dx.doi.org/10.1007/JHEP10(2021)088}{{\em JHEP} {\bfseries 10}
  (2021) 088}, \href{http://arxiv.org/abs/2106.11260}{{\ttfamily
  arXiv:2106.11260 [hep-ph]}}.

\bibitem{Chen:2022cgv}
X.~Chen, T.~Gehrmann, E.~W.~N. Glover, A.~Huss, P.~F. Monni, E.~Re, L.~Rottoli,
  and P.~Torrielli, ``{Third-Order Fiducial Predictions for Drell-Yan
  Production at the LHC},''
  \href{http://dx.doi.org/10.1103/PhysRevLett.128.252001}{{\em Phys. Rev.
  Lett.} {\bfseries 128} no.~25, (2022) 252001},
  \href{http://arxiv.org/abs/2203.01565}{{\ttfamily arXiv:2203.01565
  [hep-ph]}}.

\bibitem{Neumann:2022lft}
T.~Neumann and J.~Campbell, ``{Fiducial Drell-Yan production at the LHC
  improved by transverse-momentum resummation at N4LLp+N3LO},''
  \href{http://dx.doi.org/10.1103/PhysRevD.107.L011506}{{\em Phys. Rev. D}
  {\bfseries 107} no.~1, (2023) L011506},
  \href{http://arxiv.org/abs/2207.07056}{{\ttfamily arXiv:2207.07056
  [hep-ph]}}.

\bibitem{Collins:1984kg}
J.~C. Collins, D.~E. Soper, and G.~F. Sterman, ``{Transverse Momentum
  Distribution in Drell-Yan Pair and W and Z Boson Production},''
  \href{http://dx.doi.org/10.1016/0550-3213(85)90479-1}{{\em Nucl. Phys. B}
  {\bfseries 250} (1985) 199--224}.

\bibitem{Davies:1984sp}
C.~T.~H. Davies, B.~R. Webber, and W.~J. Stirling, ``{Drell-Yan Cross-Sections
  at Small Transverse Momentum},''.

\bibitem{Ladinsky:1993zn}
G.~A. Ladinsky and C.~P. Yuan, ``{The Nonperturbative regime in QCD resummation
  for gauge boson production at hadron colliders},''
  \href{http://dx.doi.org/10.1103/PhysRevD.50.R4239}{{\em Phys. Rev. D}
  {\bfseries 50} (1994) R4239},
  \href{http://arxiv.org/abs/hep-ph/9311341}{{\ttfamily arXiv:hep-ph/9311341}}.

\bibitem{Ellis:1997sc}
R.~K. Ellis, D.~A. Ross, and S.~Veseli, ``{Vector boson production in hadronic
  collisions},'' \href{http://dx.doi.org/10.1016/S0550-3213(97)00403-3}{{\em
  Nucl. Phys. B} {\bfseries 503} (1997) 309--338},
  \href{http://arxiv.org/abs/hep-ph/9704239}{{\ttfamily arXiv:hep-ph/9704239}}.

\bibitem{Landry:1999an}
F.~Landry, R.~Brock, G.~Ladinsky, and C.~P. Yuan, ``{New fits for the
  nonperturbative parameters in the CSS resummation formalism},''
  \href{http://dx.doi.org/10.1103/PhysRevD.63.013004}{{\em Phys. Rev. D}
  {\bfseries 63} (2000) 013004},
  \href{http://arxiv.org/abs/hep-ph/9905391}{{\ttfamily arXiv:hep-ph/9905391}}.

\bibitem{Qiu:2000hf}
J.-w. Qiu and X.-f. Zhang, ``{Role of the nonperturbative input in QCD resummed
  Drell-Yan $Q_{T}$ distributions},''
  \href{http://dx.doi.org/10.1103/PhysRevD.63.114011}{{\em Phys. Rev. D}
  {\bfseries 63} (2001) 114011},
  \href{http://arxiv.org/abs/hep-ph/0012348}{{\ttfamily arXiv:hep-ph/0012348}}.

\bibitem{Kulesza:2002rh}
A.~Kulesza, G.~F. Sterman, and W.~Vogelsang, ``{Joint resummation in
  electroweak boson production},''
  \href{http://dx.doi.org/10.1103/PhysRevD.66.014011}{{\em Phys. Rev. D}
  {\bfseries 66} (2002) 014011},
  \href{http://arxiv.org/abs/hep-ph/0202251}{{\ttfamily arXiv:hep-ph/0202251}}.

\bibitem{Konychev:2005iy}
A.~V. Konychev and P.~M. Nadolsky, ``{Universality of the Collins-Soper-Sterman
  nonperturbative function in gauge boson production},''
  \href{http://dx.doi.org/10.1016/j.physletb.2005.12.063}{{\em Phys. Lett. B}
  {\bfseries 633} (2006) 710--714},
  \href{http://arxiv.org/abs/hep-ph/0506225}{{\ttfamily arXiv:hep-ph/0506225}}.

\bibitem{Guzzi:2013aja}
M.~Guzzi, P.~M. Nadolsky, and B.~Wang, ``{Nonperturbative contributions to a
  resummed leptonic angular distribution in inclusive neutral vector boson
  production},'' \href{http://dx.doi.org/10.1103/PhysRevD.90.014030}{{\em Phys.
  Rev. D} {\bfseries 90} no.~1, (2014) 014030},
  \href{http://arxiv.org/abs/1309.1393}{{\ttfamily arXiv:1309.1393 [hep-ph]}}.

\bibitem{Collins:2014jpa}
J.~Collins and T.~Rogers, ``{Understanding the large-distance behavior of
  transverse-momentum-dependent parton densities and the Collins-Soper
  evolution kernel},'' \href{http://dx.doi.org/10.1103/PhysRevD.91.074020}{{\em
  Phys. Rev. D} {\bfseries 91} no.~7, (2015) 074020},
  \href{http://arxiv.org/abs/1412.3820}{{\ttfamily arXiv:1412.3820 [hep-ph]}}.

\bibitem{Wei:2020glg}
S.-y. Wei, ``{Exploring the non-perturbative Sudakov factor via $Z^0$ -boson
  production in $pp$ collisions},''
  \href{http://dx.doi.org/10.1016/j.physletb.2021.136356}{{\em Phys. Lett. B}
  {\bfseries 817} (2021) 136356},
  \href{http://arxiv.org/abs/2009.06514}{{\ttfamily arXiv:2009.06514
  [hep-ph]}}.

\bibitem{CDF:2012brb}
{\bfseries CDF} Collaboration, T.~Aaltonen {\em et~al.}, ``{Transverse momentum
  cross section of $e^+e^-$ pairs in the $Z$-boson region from $p\bar{p}$
  collisions at $\sqrt{s}=1.96$ TeV},''
  \href{http://dx.doi.org/10.1103/PhysRevD.86.052010}{{\em Phys. Rev. D}
  {\bfseries 86} (2012) 052010},
  \href{http://arxiv.org/abs/1207.7138}{{\ttfamily arXiv:1207.7138 [hep-ex]}}.

\bibitem{CDF:2011ksg}
{\bfseries CDF} Collaboration, T.~Aaltonen {\em et~al.}, ``{First Measurement
  of the Angular Coefficients of Drell-Yan $e^{+}e^{-}$ pairs in the Z Mass
  Region from $p\bar{p}$ Collisions at $\sqrt{s}$ = 1.96 TeV},''
  \href{http://dx.doi.org/10.1103/PhysRevLett.106.241801}{{\em Phys. Rev.
  Lett.} {\bfseries 106} (2011) 241801},
  \href{http://arxiv.org/abs/1103.5699}{{\ttfamily arXiv:1103.5699 [hep-ex]}}.

\bibitem{Camarda:2019zyx}
S.~Camarda {\em et~al.}, ``{DYTurbo: Fast predictions for Drell-Yan
  processes},'' \href{http://dx.doi.org/10.1140/epjc/s10052-020-7757-5}{{\em
  Eur. Phys. J. C} {\bfseries 80} no.~3, (2020) 251},
  \href{http://arxiv.org/abs/1910.07049}{{\ttfamily arXiv:1910.07049
  [hep-ph]}}. [Erratum: Eur.Phys.J.C 80, 440 (2020)].

\bibitem{Bozzi:2005wk}
G.~Bozzi, S.~Catani, D.~de~Florian, and M.~Grazzini, ``{Transverse-momentum
  resummation and the spectrum of the Higgs boson at the LHC},''
  \href{http://dx.doi.org/10.1016/j.nuclphysb.2005.12.022}{{\em Nucl. Phys. B}
  {\bfseries 737} (2006) 73--120},
  \href{http://arxiv.org/abs/hep-ph/0508068}{{\ttfamily arXiv:hep-ph/0508068}}.

\bibitem{Bozzi:2010xn}
G.~Bozzi, S.~Catani, G.~Ferrera, D.~de~Florian, and M.~Grazzini, ``{Production
  of Drell-Yan lepton pairs in hadron collisions: Transverse-momentum
  resummation at next-to-next-to-leading logarithmic accuracy},''
  \href{http://dx.doi.org/10.1016/j.physletb.2010.12.024}{{\em Phys. Lett. B}
  {\bfseries 696} (2011) 207--213},
  \href{http://arxiv.org/abs/1007.2351}{{\ttfamily arXiv:1007.2351 [hep-ph]}}.

\bibitem{Catani:2015vma}
S.~Catani, D.~de~Florian, G.~Ferrera, and M.~Grazzini, ``{Vector boson
  production at hadron colliders: transverse-momentum resummation and leptonic
  decay},'' \href{http://dx.doi.org/10.1007/JHEP12(2015)047}{{\em JHEP}
  {\bfseries 12} (2015) 047}, \href{http://arxiv.org/abs/1507.06937}{{\ttfamily
  arXiv:1507.06937 [hep-ph]}}.

\bibitem{Catani:2000vq}
S.~Catani, D.~de~Florian, and M.~Grazzini, ``{Universality of nonleading
  logarithmic contributions in transverse momentum distributions},''
  \href{http://dx.doi.org/10.1016/S0550-3213(00)00617-9}{{\em Nucl. Phys. B}
  {\bfseries 596} (2001) 299--312},
  \href{http://arxiv.org/abs/hep-ph/0008184}{{\ttfamily arXiv:hep-ph/0008184}}.

\bibitem{Catani:2013tia}
S.~Catani, L.~Cieri, D.~de~Florian, G.~Ferrera, and M.~Grazzini,
  ``{Universality of transverse-momentum resummation and hard factors at the
  NNLO},'' \href{http://dx.doi.org/10.1016/j.nuclphysb.2014.02.011}{{\em Nucl.
  Phys. B} {\bfseries 881} (2014) 414--443},
  \href{http://arxiv.org/abs/1311.1654}{{\ttfamily arXiv:1311.1654 [hep-ph]}}.

\bibitem{Collins:1981va}
J.~C. Collins and D.~E. Soper, ``{Back-To-Back Jets: Fourier Transform from B
  to K-Transverse},''
  \href{http://dx.doi.org/10.1016/0550-3213(82)90453-9}{{\em Nucl. Phys. B}
  {\bfseries 197} (1982) 446--476}.

\bibitem{Catani:1996yz}
S.~Catani, M.~L. Mangano, P.~Nason, and L.~Trentadue, ``{The Resummation of
  soft gluons in hadronic collisions},''
  \href{http://dx.doi.org/10.1016/0550-3213(96)00399-9}{{\em Nucl. Phys. B}
  {\bfseries 478} (1996) 273--310},
  \href{http://arxiv.org/abs/hep-ph/9604351}{{\ttfamily arXiv:hep-ph/9604351}}.

\bibitem{Laenen:2000de}
E.~Laenen, G.~F. Sterman, and W.~Vogelsang, ``{Higher order QCD corrections in
  prompt photon production},''
  \href{http://dx.doi.org/10.1103/PhysRevLett.84.4296}{{\em Phys. Rev. Lett.}
  {\bfseries 84} (2000) 4296--4299},
  \href{http://arxiv.org/abs/hep-ph/0002078}{{\ttfamily arXiv:hep-ph/0002078}}.

\bibitem{FerrarioRavasio:2020guj}
S.~Ferrario~Ravasio, G.~Limatola, and P.~Nason, ``{Infrared renormalons in
  kinematic distributions for hadron collider processes},''
  \href{http://dx.doi.org/10.1007/JHEP06(2021)018}{{\em JHEP} {\bfseries 06}
  (2021) 018}, \href{http://arxiv.org/abs/2011.14114}{{\ttfamily
  arXiv:2011.14114 [hep-ph]}}.

\bibitem{Caola:2021kzt}
F.~Caola, S.~Ferrario~Ravasio, G.~Limatola, K.~Melnikov, and P.~Nason, ``{On
  linear power corrections in certain collider observables},''
  \href{http://dx.doi.org/10.1007/JHEP01(2022)093}{{\em JHEP} {\bfseries 01}
  (2022) 093}, \href{http://arxiv.org/abs/2108.08897}{{\ttfamily
  arXiv:2108.08897 [hep-ph]}}.

\bibitem{Tafat:2001in}
S.~Tafat, ``{Nonperturbative corrections to the Drell-Yan transverse momentum
  distribution},'' \href{http://dx.doi.org/10.1088/1126-6708/2001/05/004}{{\em
  JHEP} {\bfseries 05} (2001) 004},
  \href{http://arxiv.org/abs/hep-ph/0102237}{{\ttfamily arXiv:hep-ph/0102237}}.

\bibitem{Vladimirov:2020umg}
A.~A. Vladimirov, ``{Self-contained definition of the Collins-Soper kernel},''
  \href{http://dx.doi.org/10.1103/PhysRevLett.125.192002}{{\em Phys. Rev.
  Lett.} {\bfseries 125} no.~19, (2020) 192002},
  \href{http://arxiv.org/abs/2003.02288}{{\ttfamily arXiv:2003.02288
  [hep-ph]}}.

\bibitem{Schweitzer:2012hh}
P.~Schweitzer, M.~Strikman, and C.~Weiss, ``{Intrinsic transverse momentum and
  parton correlations from dynamical chiral symmetry breaking},''
  \href{http://dx.doi.org/10.1007/JHEP01(2013)163}{{\em JHEP} {\bfseries 01}
  (2013) 163}, \href{http://arxiv.org/abs/1210.1267}{{\ttfamily arXiv:1210.1267
  [hep-ph]}}.

\bibitem{Scimemi:2019cmh}
I.~Scimemi and A.~Vladimirov, ``{Non-perturbative structure of semi-inclusive
  deep-inelastic and Drell-Yan scattering at small transverse momentum},''
  \href{http://dx.doi.org/10.1007/JHEP06(2020)137}{{\em JHEP} {\bfseries 06}
  (2020) 137}, \href{http://arxiv.org/abs/1912.06532}{{\ttfamily
  arXiv:1912.06532 [hep-ph]}}.

\bibitem{Boughezal:2015ded}
R.~Boughezal, J.~M. Campbell, R.~K. Ellis, C.~Focke, W.~T. Giele, X.~Liu, and
  F.~Petriello, ``{Z-boson production in association with a jet at
  next-to-next-to-leading order in perturbative QCD},''
  \href{http://dx.doi.org/10.1103/PhysRevLett.116.152001}{{\em Phys. Rev.
  Lett.} {\bfseries 116} no.~15, (2016) 152001},
  \href{http://arxiv.org/abs/1512.01291}{{\ttfamily arXiv:1512.01291
  [hep-ph]}}.

\bibitem{Camarda:2021jsw}
S.~Camarda, L.~Cieri, and G.~Ferrera, ``{Fiducial perturbative power
  corrections within the $\mathbf{q}_T$ subtraction formalism},''
  \href{http://dx.doi.org/10.1140/epjc/s10052-022-10510-x}{{\em Eur. Phys. J.
  C} {\bfseries 82} no.~6, (2022) 575},
  \href{http://arxiv.org/abs/2111.14509}{{\ttfamily arXiv:2111.14509
  [hep-ph]}}.

\bibitem{Billis:2021ecs}
G.~Billis, B.~Dehnadi, M.~A. Ebert, J.~K.~L. Michel, and F.~J. Tackmann,
  ``{Higgs pT Spectrum and Total Cross Section with Fiducial Cuts at Third
  Resummed and Fixed Order in QCD},''
  \href{http://dx.doi.org/10.1103/PhysRevLett.127.072001}{{\em Phys. Rev.
  Lett.} {\bfseries 127} no.~7, (2021) 072001},
  \href{http://arxiv.org/abs/2102.08039}{{\ttfamily arXiv:2102.08039
  [hep-ph]}}.

\bibitem{vanRitbergen:1997va}
T.~van Ritbergen, J.~A.~M. Vermaseren, and S.~A. Larin, ``{The Four loop beta
  function in quantum chromodynamics},''
  \href{http://dx.doi.org/10.1016/S0370-2693(97)00370-5}{{\em Phys. Lett. B}
  {\bfseries 400} (1997) 379--384},
  \href{http://arxiv.org/abs/hep-ph/9701390}{{\ttfamily arXiv:hep-ph/9701390}}.

\bibitem{Czakon:2004bu}
M.~Czakon, ``{The Four-loop QCD beta-function and anomalous dimensions},''
  \href{http://dx.doi.org/10.1016/j.nuclphysb.2005.01.012}{{\em Nucl. Phys. B}
  {\bfseries 710} (2005) 485--498},
  \href{http://arxiv.org/abs/hep-ph/0411261}{{\ttfamily arXiv:hep-ph/0411261}}.

\bibitem{Buckley:2014ana}
A.~Buckley, J.~Ferrando, S.~Lloyd, K.~Nordstr\"om, B.~Page, M.~R\"ufenacht,
  M.~Sch\"onherr, and G.~Watt, ``{LHAPDF6: parton density access in the LHC
  precision era},''
  \href{http://dx.doi.org/10.1140/epjc/s10052-015-3318-8}{{\em Eur. Phys. J. C}
  {\bfseries 75} (2015) 132}, \href{http://arxiv.org/abs/1412.7420}{{\ttfamily
  arXiv:1412.7420 [hep-ph]}}.

\bibitem{Vogt:2004ns}
A.~Vogt, ``{Efficient evolution of unpolarized and polarized parton
  distributions with QCD-PEGASUS},''
  \href{http://dx.doi.org/10.1016/j.cpc.2005.03.103}{{\em Comput. Phys.
  Commun.} {\bfseries 170} (2005) 65--92},
  \href{http://arxiv.org/abs/hep-ph/0408244}{{\ttfamily arXiv:hep-ph/0408244}}.

\bibitem{Alekhin:2014irh}
S.~Alekhin {\em et~al.}, ``{HERAFitter},''
  \href{http://dx.doi.org/10.1140/epjc/s10052-015-3480-z}{{\em Eur. Phys. J. C}
  {\bfseries 75} no.~7, (2015) 304},
  \href{http://arxiv.org/abs/1410.4412}{{\ttfamily arXiv:1410.4412 [hep-ph]}}.

\bibitem{HERAFitterdevelopersTeam:2015cre}
{\bfseries HERAFitter developers' Team} Collaboration, S.~Camarda {\em et~al.},
  ``{QCD analysis of $W$- and $Z$-boson production at Tevatron},''
  \href{http://dx.doi.org/10.1140/epjc/s10052-015-3655-7}{{\em Eur. Phys. J. C}
  {\bfseries 75} no.~9, (2015) 458},
  \href{http://arxiv.org/abs/1503.05221}{{\ttfamily arXiv:1503.05221
  [hep-ph]}}.

\bibitem{Sjostrand:2014zea}
T.~Sj\"ostrand, S.~Ask, J.~R. Christiansen, R.~Corke, N.~Desai, P.~Ilten,
  S.~Mrenna, S.~Prestel, C.~O. Rasmussen, and P.~Z. Skands, ``{An introduction
  to PYTHIA 8.2}'' \href{http://dx.doi.org/10.1016/j.cpc.2015.01.024}{{\em
  Comput. Phys. Commun.} {\bfseries 191} (2015) 159--177},
  \href{http://arxiv.org/abs/1410.3012}{{\ttfamily arXiv:1410.3012 [hep-ph]}}.

\bibitem{Cieri:2018sfk}
L.~Cieri, G.~Ferrera, and G.~F.~R. Sborlini, ``{Combining QED and QCD
  transverse-momentum resummation for Z boson production at hadron
  colliders},'' \href{http://dx.doi.org/10.1007/JHEP08(2018)165}{{\em JHEP}
  {\bfseries 08} (2018) 165}, \href{http://arxiv.org/abs/1805.11948}{{\ttfamily
  arXiv:1805.11948 [hep-ph]}}.

\bibitem{Carrazza:2015aoa}
S.~Carrazza, S.~Forte, Z.~Kassabov, J.~I. Latorre, and J.~Rojo, ``{An Unbiased
  Hessian Representation for Monte Carlo PDFs},''
  \href{http://dx.doi.org/10.1140/epjc/s10052-015-3590-7}{{\em Eur. Phys. J. C}
  {\bfseries 75} no.~8, (2015) 369},
  \href{http://arxiv.org/abs/1505.06736}{{\ttfamily arXiv:1505.06736
  [hep-ph]}}.

\bibitem{NNPDF:2021njg}
{\bfseries NNPDF} Collaboration, R.~D. Ball {\em et~al.}, ``{The path to proton
  structure at 1\% accuracy},''
  \href{http://dx.doi.org/10.1140/epjc/s10052-022-10328-7}{{\em Eur. Phys. J.
  C} {\bfseries 82} no.~5, (2022) 428},
  \href{http://arxiv.org/abs/2109.02653}{{\ttfamily arXiv:2109.02653
  [hep-ph]}}.

\bibitem{Bailey:2020ooq}
S.~Bailey, T.~Cridge, L.~A. Harland-Lang, A.~D. Martin, and R.~S. Thorne,
  ``{Parton distributions from LHC, HERA, Tevatron and fixed target data:
  MSHT20 PDFs},'' \href{http://dx.doi.org/10.1140/epjc/s10052-021-09057-0}{{\em
  Eur. Phys. J. C} {\bfseries 81} no.~4, (2021) 341},
  \href{http://arxiv.org/abs/2012.04684}{{\ttfamily arXiv:2012.04684
  [hep-ph]}}.

\bibitem{H1:2015ubc}
{\bfseries H1, ZEUS} Collaboration, H.~Abramowicz {\em et~al.}, ``{Combination
  of measurements of inclusive deep inelastic ${e^{\pm }p}$ scattering cross
  sections and QCD analysis of HERA data},''
  \href{http://dx.doi.org/10.1140/epjc/s10052-015-3710-4}{{\em Eur. Phys. J. C}
  {\bfseries 75} no.~12, (2015) 580},
  \href{http://arxiv.org/abs/1506.06042}{{\ttfamily arXiv:1506.06042
  [hep-ex]}}.

\bibitem{McGowan:2022nag}
J.~McGowan, T.~Cridge, L.~A. Harland-Lang, and R.~S. Thorne, ``{Approximate
  N$^{3}$LO parton distribution functions with theoretical uncertainties:
  MSHT20aN$^3$LO PDFs},''
  \href{http://dx.doi.org/10.1140/epjc/s10052-023-11236-0}{{\em Eur. Phys. J.
  C} {\bfseries 83} no.~3, (2023) 185},
  \href{http://arxiv.org/abs/2207.04739}{{\ttfamily arXiv:2207.04739
  [hep-ph]}}. [Erratum: Eur.Phys.J.C 83, 302 (2023)].

\bibitem{Camarda:2023dqn}
S.~Camarda, L.~Cieri, and G.~Ferrera, ``{Drell-Yan lepton-pair production:
  $q_T$ resummation at approximate N$^4$LL+N$^4$LO accuracy},''
  \href{http://arxiv.org/abs/2303.12781}{{\ttfamily arXiv:2303.12781
  [hep-ph]}}.

\bibitem{Forte:2020pyp}
S.~Forte and Z.~Kassabov, ``{Why $\alpha _s$ cannot be determined from hadronic
  processes without simultaneously determining the parton distributions},''
  \href{http://dx.doi.org/10.1140/epjc/s10052-020-7748-6}{{\em Eur. Phys. J. C}
  {\bfseries 80} no.~3, (2020) 182},
  \href{http://arxiv.org/abs/2001.04986}{{\ttfamily arXiv:2001.04986
  [hep-ph]}}.

\bibitem{Paukkunen:2014zia}
H.~Paukkunen and P.~Zurita, ``{PDF reweighting in the Hessian matrix
  approach},'' \href{http://dx.doi.org/10.1007/JHEP12(2014)100}{{\em JHEP}
  {\bfseries 12} (2014) 100}, \href{http://arxiv.org/abs/1402.6623}{{\ttfamily
  arXiv:1402.6623 [hep-ph]}}.

\bibitem{Weinberg:1973xwm}
S.~Weinberg, ``{New approach to the renormalization group},''
  \href{http://dx.doi.org/10.1103/PhysRevD.8.3497}{{\em Phys. Rev. D}
  {\bfseries 8} (1973) 3497--3509}.

\bibitem{Thorne:1997ga}
R.~S. Thorne and R.~G. Roberts, ``Ordered analysis of heavy flavor production
  in deep-inelastic scattering,''
  \href{http://dx.doi.org/10.1103/PhysRevD.57.6871}{{\em Phys. Rev. D}
  {\bfseries 57} (1998) 6871--6898},
  \href{http://arxiv.org/abs/hep-ph/9709442}{{\ttfamily arXiv:hep-ph/9709442}}.

\bibitem{Thorne:2006qt}
R.~S. Thorne, ``Variable-flavor number scheme for next-to-next-to-leading
  order,'' \href{http://dx.doi.org/10.1103/PhysRevD.73.054019}{{\em Phys. Rev.
  D} {\bfseries 73} (2006) 054019},
  \href{http://arxiv.org/abs/hep-ph/0601245}{{\ttfamily arXiv:hep-ph/0601245}}.

\bibitem{Thorne:2012az}
R.~S. Thorne, ``{Effect of changes of variable flavor number scheme on parton
  distribution functions and predicted cross sections},''
  \href{http://dx.doi.org/10.1103/PhysRevD.86.074017}{{\em Phys. Rev. D}
  {\bfseries 86} (2012) 074017},
  \href{http://arxiv.org/abs/1201.6180}{{\ttfamily arXiv:1201.6180 [hep-ph]}}.

\bibitem{Bacchetta:2019sam}
A.~Bacchetta, V.~Bertone, C.~Bissolotti, G.~Bozzi, F.~Delcarro, F.~Piacenza,
  and M.~Radici, ``{Transverse-momentum-dependent parton distributions up to
  N$^{3}$LL from Drell-Yan data},''
  \href{http://dx.doi.org/10.1007/JHEP07(2020)117}{{\em JHEP} {\bfseries 07}
  (2020) 117}, \href{http://arxiv.org/abs/1912.07550}{{\ttfamily
  arXiv:1912.07550 [hep-ph]}}.

\bibitem{Sun:2014dqm}
P.~Sun, J.~Isaacson, C.~P. Yuan, and F.~Yuan, ``{Nonperturbative functions for
  SIDIS and Drell\textendash{}Yan processes},''
  \href{http://dx.doi.org/10.1142/S0217751X18410063}{{\em Int. J. Mod. Phys. A}
  {\bfseries 33} no.~11, (2018) 1841006},
  \href{http://arxiv.org/abs/1406.3073}{{\ttfamily arXiv:1406.3073 [hep-ph]}}.

\bibitem{Parisi:1979se}
G.~Parisi and R.~Petronzio, ``{Small Transverse Momentum Distributions in Hard
  Processes},'' \href{http://dx.doi.org/10.1016/0550-3213(79)90040-3}{{\em
  Nucl. Phys. B} {\bfseries 154} (1979) 427--440}.

\bibitem{Hirai:2012ada}
M.~Hirai, H.~Kawamura, and K.~Tanaka,
  \href{http://dx.doi.org/10.3204/DESY-PROC-2012-02/136}{``{New determination
  of the nonperturbative form factor in QCD transverse-momentum resummation for
  vector boson production},''} in {\em {20th International Workshop on
  Deep-Inelastic Scattering and Related Subjects}}, pp.~535--538.
\newblock 2012.

\bibitem{D0:2010qhp}
{\bfseries D0} Collaboration, V.~M. Abazov {\em et~al.}, ``{Precise Study of
  the $Z/\gamma^*$ Boson Transverse Momentum Distribution in $p\bar{p}$
  Collisions using a Novel Technique},''
  \href{http://dx.doi.org/10.1103/PhysRevLett.106.122001}{{\em Phys. Rev.
  Lett.} {\bfseries 106} (2011) 122001},
  \href{http://arxiv.org/abs/1010.0262}{{\ttfamily arXiv:1010.0262 [hep-ex]}}.

\bibitem{Bethke:1992jn}
S.~Bethke and S.~Catani, ``{A Summary of alpha-s measurements},'' in {\em {27th
  Rencontres de Moriond: QCD and High-energy Hadronic Interactions}},
  pp.~203--208.
\newblock 1992.

\bibitem{Catani:1992ua}
S.~Catani, L.~Trentadue, G.~Turnock, and B.~R. Webber, ``{Resummation of large
  logarithms in e+ e- event shape distributions},''
  \href{http://dx.doi.org/10.1016/0550-3213(93)90271-P}{{\em Nucl. Phys. B}
  {\bfseries 407} (1993) 3--42}.

\bibitem{Verbytskyi:2019zhh}
A.~Verbytskyi, A.~Banfi, A.~Kardos, P.~F. Monni, S.~Kluth, G.~Somogyi,
  Z.~Sz\H{o}r, Z.~Tr\'ocs\'anyi, Z.~Tulip\'ant, and G.~Zanderighi, ``{High
  precision determination of $\alpha_s$ from a global fit of jet rates},''
  \href{http://dx.doi.org/10.1007/JHEP08(2019)129}{{\em JHEP} {\bfseries 08}
  (2019) 129}, \href{http://arxiv.org/abs/1902.08158}{{\ttfamily
  arXiv:1902.08158 [hep-ph]}}.

\end{thebibliography}\endgroup

\clearpage
\appendix
\section{Matching corrections}
\label{app:match}
In this Appendix we discuss the interpolation of the $\textrm{d}\sigma^{\textrm{f.o.}}-\textrm{d}\sigma^{\textrm{asy}}$
matching corrections of Eq.~(\ref{eq:rescross_1}) at $\mathcal{O}(\as^3)$ with their expected quadratic dependence on
$\pt/m_Z$ using the function
\begin{eqnarray}
  \frac{\pt^2}{m_Z^2}\sum_i c_i \ln^i\left(\frac{\pt}{m_Z}\right).
\end{eqnarray}

Fits are performed in the region of $\pt < 50$~GeV, with 10
logarithmically spaced bins.
The p-values for fits of the matching corrections with different
choices of the renormalisation, factorisation, and resummation scales
are in the range from $0.3$ to $0.9$.

We have considered two sources of uncertainties
addressing the statistical and systematic uncertainties of the
matching corrections.
We have varied the lower cutoff from $p_T = 5$~GeV to $p_T =
10$~GeV. The difference in $\alpha_S(m_Z)$ of $0.0001$ is considered as
an additional systematic uncertainty.
In order to estimate the statistical uncertainty, we have generated
a set of 1000 Monte Carlo replicas of the matching corrections, by
fluctuating them within their numerical uncertainties. The upper and
lower limits of the 68$\%$ confidence level envelope of the
extrapolation fits to the 1000 replicas are used for the estimate of
the statistical uncertainty, yielding $\pm 0.00002$ on $\alpha_S(m_Z)$.

\begin{figure}
  \begin{center}
    \includegraphics[width=0.45\textwidth]{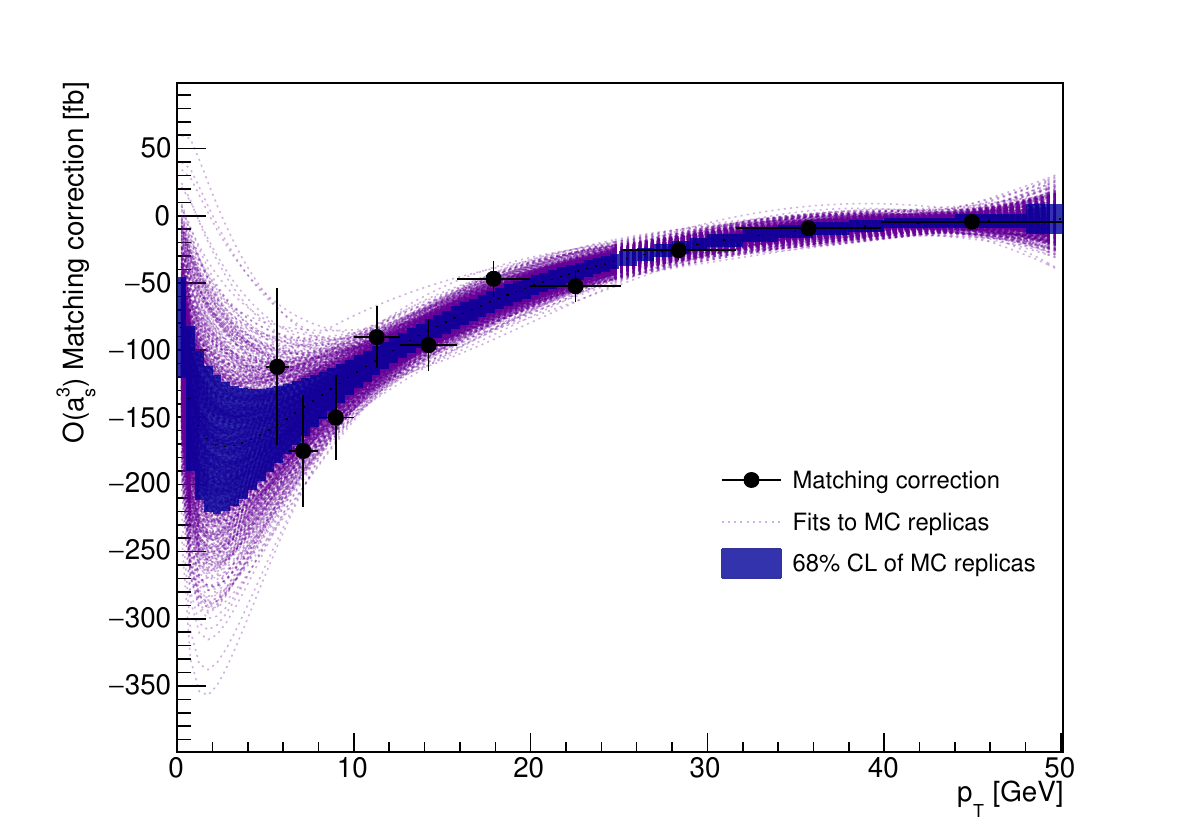}
  \end{center}
  \caption{
    \label{fig:match}
    The $\mathcal{O}(\as^3)$ matching corrections }
\end{figure}

The difference between the NNLO $Z$+jet predictions and the expansion of the resummed calculation,
showing the numerical accuracy in the matching procedure at $\alpha_S^3$ order, is presented in Fig.~\ref{fig:match},
which also shows the replicas and their 68$\%$ confidence level uncertainty band.

Comparable figures showing the difference of the
asymptotic term $\textrm{d}\sigma^{\textrm{asy}}$ and the \Vboson+jet
finite-order cross section $\textrm{d}\sigma^{\textrm{f.o.}}$ at
$\mathcal{O}(\as^3)$ can be found in Refs.~\cite{Camarda:2021ict,Chen:2022cgv,Neumann:2022lft,Camarda:2021jsw}

The studies of statistical and systematic uncertainties
discussed above confirm that the $\mathcal{O}(\as^3)$ matching corrections are associated
with small uncertainties, which are accounted for in the final
result. The estimated uncertainties of $\pm0.0001$ are consistent
with the overall small impact of such corrections, which is
estimated as $+0.0005$.
  
\section{Simultaneous PDF and \asmz{} fit}
\label{app:pdffit}
The Hessian profiling employed in this analysis provides an
approximation to a PDF determination which relies on the accuracy of
the quadratic approximation around the minimum~\cite{Paukkunen:2014zia}.
The validity of the Hessian profiling approximation, is verified by
performing a simultaneous fit of PDFs, \asmz, and the
the parameter $g$ of the Gaussian non-perturbative
form factor, with a setup similar to the one employed for the
HERAPDF2.0~\cite{H1:2015ubc} PDF determination. In this Appendix we provide further quantitative details
of the comparison of the Hessian profiling of HERAPDF2.0 with such a PDF
fit. The PDF fit includes the combined neutral and charged current
deep inelastic scattering (DIS) cross-section data from the H1 and
ZEUS experiments at the HERA collider~\cite{H1:2015ubc}.
Table~\ref{tab:fit} shows the contribution to the total $\chi^2$ of
the various datasets used in the fit, compared to the $\chi^2$ of the
Hessian profiling, and a comparison of the determined values of
\asmz{} and $g$.

\begin{table}[h]
  \begin{center}
    \caption{\label{tab:fit} Comparison of the Hessian profiling of
      HERAPDF2.0 with the PDF fit, including the contribution to the total $\chi^2$ at
      minimum of the various datasets used in the fit.}
    \begin{tabular}{lcc}
      \toprule
      & PDF fit   & Hessian \\
      & & profiling \\
      \midrule
      \asmz          & $0.1188 \pm 0.0008$ & $0.1184 \pm 0.0006$ \\
      $g$ [GeV$^2$]  & $0.69 \pm 0.05$ & $0.71 \pm 0.05$ \\
      \midrule
      Dataset       & $\chi^2$/points  & $\chi^2$/points  \\
      \midrule
  NC DIS H1-ZEUS  $e^+p$     & $955/905$  \\ 
  CC DIS H1-ZEUS  $e^+p$     & $46/39$  \\ 
  NC DIS H1-ZEUS  $e^-p$     & $219/159$  \\ 
  CC DIS H1-ZEUS  $e^-p$     & $53/42$  \\ 
  H1-ZEUS correlated $\chi^2$               & $91$  \\ 
      \midrule
  CDF $Z$ \pt{}                 & $41/55$ & $40/55$ \\ 
      \midrule
  Total   & 1405 / 1184  \\ 
      \bottomrule
    \end{tabular}
  \end{center}
\end{table}

\end{document}